\documentstyle[12pt]{article}
\topmargin - 1.0 true cm
\newcommand{\beq}{\begin{equation}}
\newcommand{\eeq}{\end{equation}}
\def\sss{\scriptscriptstyle}

\def\bd{B_d^0}
\def\bdbar{{\overline{B_d^0}}}
\def\bs{B_s^0}
\def\bsbar{{\overline{B_s^0}}}
\def\barp{{\raise.35ex\hbox{${\sss (}$}}---{\raise.35ex\hbox{${\sss )}$}}}
\def\bdbarp{\hbox{$B_d$\kern-1.4em\raise1.4ex\hbox{\barp}}}
\def\bdobarp{\hbox{$B_d^0$\kern-1.4em\raise1.4ex\hbox{\barp}}}
\def\bsbarp{\hbox{$B_s$\kern-1.4em\raise1.4ex\hbox{\barp}}}
\def\ks{K_{\sss S}}
\def\kl{K_{\sss L}}

\def\BK{B_{\sss K}}
\def\BBd{B_{\sss B_d}}
\def\BBs{B_{\sss B_s}}
\def\roughly#1{\mathrel{\raise.3ex\hbox{$#1$\kern-.75em\lower1ex\hbox{$\sim$}}}}
\def\lsim{\roughly<}

\def\twida{\tilde{\alpha}}
\def\twidb{\tilde{\beta}}
\def\twidg{\tilde{\gamma}}

\def\epjc#1#2#3{{\it Eur.\ Phys.\ J.}\ {\bf C#1}, #3 (19#2)}

\def\npb#1#2#3{{\it Nucl.\ Phys.} {\bf B#1}, #3 (19#2)}
\def\plb#1#2#3{{\it Phys.\ Lett.} {\bf #1B}, #3 (19#2)}
\def\prd#1#2#3{{\it Phys.\ Rev.} {\bf D#1}, #3 (19#2)}
\def\newprd#1#2#3{{\it Phys.\ Rev.} {\bf D#1}: #3 (19#2)}

\def\prl#1#2#3{{\it Phys.\ Rev.\ Lett.} {\bf #1}, #3 (19#2)}
\def\zpc#1#2#3{{\it Zeit.\ Phys.} {\bf C#1}, #3 (19#2)}
\def\ijmp#1#2#3{{\it Int.\ J.\ Mod.\ Phys.} {\bf A#1}, #3 (19#2)}

\newread\epsffilein % file to \read
\newif\ifepsffileok % continue looking for the bounding box?
\newif\ifepsfbbfound % success?
\newif\ifepsfverbose % report what you're making?
\newdimen\epsfxsize % horizontal size after scaling
\newdimen\epsfysize % vertical size after scaling
\newdimen\epsftsize % horizontal size before scaling
\newdimen\epsfrsize % vertical size before scaling
\newdimen\epsftmp % register for arithmetic manipulation
\newdimen\pspoints % conversion factor
\def\epsfbox#1{\global\def\epsfllx{72}\global\def\epsflly{72}%
\global\def\epsfurx{540}\global\def\epsfury{720}%
\def\lbracket{[}\def\testit{#1}\ifx\testit\lbracket
\let\next=\epsfgetlitbb\else\let\next=\epsfnormal\fi\next{#1}}%
\def\epsfgetlitbb#1#2 #3 #4 #5]#6{\epsfgrab #2 #3 #4 #5 .\\%
\epsfsetgraph{#6}}%
\def\epsfnormal#1{\epsfgetbb{#1}\epsfsetgraph{#1}}%
\def\epsfgetbb#1{%
\openin\epsffilein=#1
\ifeof\epsffilein\errmessage{I couldn't open #1, will ignore it}\else
{\epsffileoktrue \chardef\other=12
\def\do##1{\catcode`##1=\other}\dospecials \catcode`\ =10
\loop
\read\epsffilein to \epsffileline
\ifeof\epsffilein\epsffileokfalse\else
\expandafter\epsfaux\epsffileline:. \\%
\fi
\ifepsffileok\repeat
\ifepsfbbfound\else
\ifepsfverbose\message{No bounding box comment in #1; using defaults}\fi\fi
}\closein\epsffilein\fi}%
\def\epsfclipstring{}% do we clip or not? If so,
\def\epsfsetgraph#1{%
\epsfrsize=\epsfury\pspoints
\advance\epsfrsize by-\epsflly\pspoints
\epsftsize=\epsfurx\pspoints
\advance\epsftsize by-\epsfllx\pspoints
\epsfxsize\epsfsize\epsftsize\epsfrsize
\ifnum\epsfxsize=0 \ifnum\epsfysize=0
\epsfxsize=\epsftsize \epsfysize=\epsfrsize
\epsfrsize=0pt
\else\epsftmp=\epsftsize \divide\epsftmp\epsfrsize
\epsfxsize=\epsfysize \multiply\epsfxsize\epsftmp
\multiply\epsftmp\epsfrsize \advance\epsftsize-\epsftmp
\epsftmp=\epsfysize
\loop \advance\epsftsize\epsftsize \divide\epsftmp 2
\ifnum\epsftmp>0
\ifnum\epsftsize<\epsfrsize\else
\advance\epsftsize-\epsfrsize \advance\epsfxsize\epsftmp \fi
\repeat
\epsfrsize=0pt
\fi
\else \ifnum\epsfysize=0
\epsftmp=\epsfrsize \divide\epsftmp\epsftsize
\epsfysize=\epsfxsize \multiply\epsfysize\epsftmp
\multiply\epsftmp\epsftsize \advance\epsfrsize-\epsftmp
\epsftmp=\epsfxsize
\loop \advance\epsfrsize\epsfrsize \divide\epsftmp 2
\ifnum\epsftmp>0
\ifnum\epsfrsize<\epsftsize\else
\advance\epsfrsize-\epsftsize \advance\epsfysize\epsftmp \fi
\repeat
\epsfrsize=0pt
\else
\epsfrsize=\epsfysize
\fi
\fi
\ifepsfverbose\message{#1: width=\the\epsfxsize, height=\the\epsfysize}\fi
\epsftmp=10\epsfxsize \divide\epsftmp\pspoints
\vbox to\epsfysize{\vfil\hbox to\epsfxsize{%
\ifnum\epsfrsize=0\relax
\includegraphics{#1}%
\else
\epsfrsize=10\epsfysize \divide\epsfrsize\pspoints
\includegraphics{#1}%
\fi
\hfil}}%
\global\epsfxsize=0pt\global\epsfysize=0pt}%
{\catcode`\%=12 
\global\let\epsfpercent=%\global\def\epsfbblit{%BoundingBox}}%
\long\def\epsfaux#1#2:#3\\{\ifx#1\epsfpercent
\def\testit{#2}\ifx\testit\epsfbblit
\epsfgrab #3 . . . \\%
\epsffileokfalse
\global\epsfbbfoundtrue
\fi\else\ifx#1\par\else\epsffileokfalse\fi\fi}%
\def\epsfempty{}%
\def\epsfgrab #1 #2 #3 #4 #5\\{%
\global\def\epsfllx{#1}\ifx\epsfllx\epsfempty
\epsfgrab #2 #3 #4 #5 .\\\else
\global\def\epsflly{#2}%
\global\def\epsfurx{#3}\global\def\epsfury{#4}\fi}%
\def\epsfsize#1#2{\epsfxsize}
\begin{document}
\baselineskip=6truemm
\begin{flushright}
NSF-PT-99-4 \\
UdeM-GPP-TH-98-48 \\
\end{flushright}
\bigskip
\begin{center}
{\Large \bf
$B$-Decay CP Asymmetries, Discrete Ambiguities}
\smallskip\\
{\Large \bf
and New Physics}
\bigskip\\
{\large Boris Kayser$^a$ and David London$^b$}
\end{center}

\bigskip

\begin{flushleft}
~~~~~~~~~~~$a$: {\it Division of Physics, National Science
  Foundation, 4201 Wilson Blvd.,}\\
~~~~~~~~~~~~~~~ {\it Arlington, VA 22230 USA}\\
~~~~~~~~~~~$b$: {\it Laboratoire Ren\'e J.-A. L\'evesque,
  Universit\'e de Montr\'eal,}\\
~~~~~~~~~~~~~~~{\it C.P. 6128, succ. centre-ville, Montr\'eal, QC,
  Canada H3C 3J7}
\end{flushleft}

\medskip

\begin{center}
(\today)
\end{center}

\bigskip

\begin{quote}
  {\bf Abstract}: The first measurements of CP violation in the $B$
  system will likely probe $\sin 2\alpha$, $\sin 2\beta$ and $\cos
  2\gamma$. Assuming that the CP angles $\alpha$, $\beta$ and $\gamma$
  are the interior angles of the unitarity triangle, these
  measurements determine the angle set $(\alpha,\beta,\gamma)$ except
  for a twofold discrete ambiguity. If one allows for the possibility
  of new physics, the presence of this discrete ambiguity can make its
  discovery difficult: if only one of the two candidate solutions is
  consistent with constraints from other measurements in the $B$ and
  $K$ systems, one is not sure whether new physics is present or not.
  We review the methods used to resolve the discrete ambiguity and
  show that, even in the presence of new physics, they can usually be
  used to uncover this new physics. There are some exceptions, which
  we describe in detail. We systematically scan the parameter space
  and present examples of values of $(\alpha,\beta,\gamma)$ and the
  new-physics parameters which correspond to all possibilities.
  Finally, we show that if one relaxes the assumption that the bag
  parameters $\BBd$ and $\BK$ are positive, one can no longer
  definitively establish the presence of new physics.
\end{quote}
\newpage

\section{Introduction}

In the Standard Model (SM), CP violation is due to nonzero complex
phases in the Cabibbo-Kobayashi-Maskawa (CKM) mixing matrix $V$. These
CKM phases are elegantly described in terms of the interior angles
$\alpha$, $\beta$ and $\gamma$ of the ``unitarity triangle''
\cite{PDG}. This triangle has two possible orientations. If we take
the possible range of any angle to be $-\pi$ to $+\pi$, then, for one
of these orientations, $\alpha$, $\beta$ and $\gamma$ are all
positive, while for the other they are all negative. Either way,
\beq
\label{tricond1}
{\hbox{\rm $\alpha$, $\beta$ and $\gamma$ are all of the same sign.}}
\eeq
In addition, the angles in the unitarity triangle obviously satisfy
the constraint
\beq
\label{tricond2}
\left\vert \alpha + \beta + \gamma \right\vert = \pi ~.
\eeq

\begin{figure}
\vskip -1.0truein
\centerline{\epsfxsize 3.5 truein \epsfbox {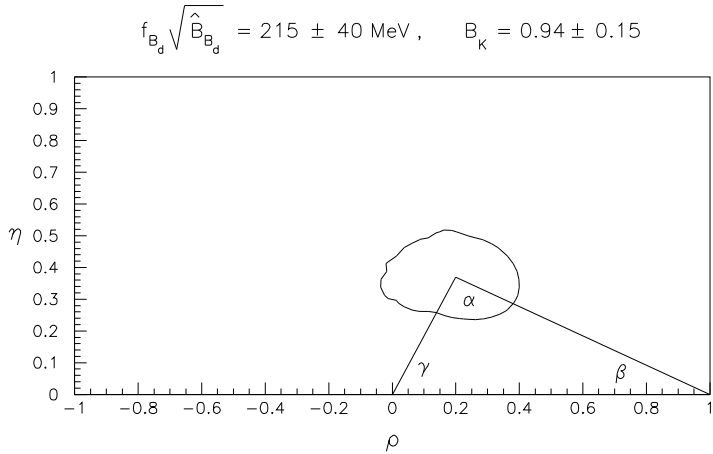}}
\vskip -1.4truein
\caption{Allowed region (95\% C.L.) in $\rho$--$\eta$ space in the SM
\protect\cite{AliLon}.}
\label{rhoeta1}
\end{figure}

The angles $\alpha$, $\beta$ and $\gamma$ may be expressed in terms of
the parameters $\rho$ and $\eta$ in Wolfenstein's approximation to the
CKM matrix \cite{Wolfenstein}. Existing information on $|V_{cb}|$,
$|V_{ub}/V_{cb}|$, $B_d$ and $B_s$ mixing, and CP violation in the
kaon system ($\epsilon_K$) restricts $\rho$ and $\eta$ to the 95\%
confidence level allowed region shown in Fig.~\ref{rhoeta1}
\cite{AliLon}. Correspondingly, $\alpha$, $\beta$ and $\gamma$ are
restricted to the ranges
\begin{eqnarray}
65^\circ \le & \alpha & \le 123^\circ ~,  \label{betaconstraint}  \\
16^\circ \le & \beta &  \le  35^\circ ~,  \label{betac2}  \\
37^\circ \le & \gamma & \le  97^\circ ~.  \label{betac3}
\end{eqnarray}
Note that the unitarity triangle shown in Fig.~\ref{rhoeta1} points
up, which implies that the CP angles $\alpha$, $\beta$ and $\gamma$
are all positive. This is a consequence of the measured phase of
$\epsilon_K$ and of the assumption that the kaon bag parameter $\BK$
is positive \cite{NirQuinn,signBK}. While lattice calculations firmly
indicate that, in fact, $\BK > 0$, this has not been verified
experimentally. If $\BK < 0$, the unitarity triangle shown in
Fig.~\ref{rhoeta1} points down, so that the CP angles are all
%
%DL: added sentences about BK.
%
negative, with the above allowed ranges changing sign as well. For the
most part, in this paper we assume, as usual, that the lattice
prediction that $\BK > 0$ is correct. However, we shall also include a
separate discussion of the consequences for the search for new physics
if it is not.

% This is Figure 1

To test the SM picture of CP violation, and to look for evidence of
new physics beyond the SM, coming experiments will attempt to
determine the angles $\alpha$, $\beta$ and $\gamma$ implied by
CP-violating asymmetries in various $B$ decays \cite{CPreview}. If the
measured angles violate either of the ``triangle conditions,''
Eqs.~(\ref{tricond1}) and (\ref{tricond2}), or correspond to a point
$(\rho,\eta)$ which is outside the allowed region
(Fig.~\ref{rhoeta1}), then we will have evidence of new physics.

The sign of the CP asymmetries in $B_d$ decays is dependent on the
sign of the bag parameter $\BBd$ \cite{signBK}. Like $\BK$, $\BBd$ is
firmly predicted by lattice calculations to be positive, and is
usually assumed to have this sign. Under this assumption, if
measurements yield a unitarity triangle which points downward, i.e.\
one which is inconsistent with the measurement of $\epsilon_K$, this
implies that new physics is present, even if the angles of the
downward-pointing triangle have magnitudes consistent with
Eqs.~(\ref{betaconstraint})-(\ref{betac3}). The CDF collaboration has
recently reported that, at the 93\% confidence level, $\Gamma
[\overline{B_d}(t) \rightarrow \Psi \ks] > \Gamma [B_d(t) \rightarrow
\Psi \ks]$ \cite{CDF99}. If we combine this result with the constraint
of Eq.~(\ref{betac2}) on $|\beta|$, and assume that $\BBd > 0$, this
result implies that the unitarity triangle points upward, consistent
with the implication of $\epsilon_K$.

Unfortunately, the CP asymmetries in the $B$ system do not directly
determine the angles in the unitarity triangle. Rather, these
asymmetries yield only trigonometric functions of these angles, such
as $\sin 2\alpha$ and $\sin 2\beta$, leaving the underlying angles
themselves discretely ambiguous. Needless to say, this makes the goal
of testing for consistency with the SM and looking for evidence of
physics beyond it more challenging.

The first CP asymmetry to be measured will almost certainly be the one
in $\bd(t) \to \Psi \ks$. The second may well be the one in $\bd(t)
\to \pi^+\pi^-$ (or $\pi^+\pi^-\pi^0$ \cite{Dalitz}). Quite possibly,
the third will be the one in $B^\pm \to D K^\pm$ \cite{BtoDK}. We
shall assume this scenario. Now, the CP asymmetry in any $B$ decay
probes the CP-odd part of the relative phase of (hopefully only) two
interfering amplitudes. We shall call the CP-odd relative phases
probed in $\bd(t) \to \pi^+\pi^-$, $\bd(t) \to \Psi \ks$ and $B^\pm
\to D K^\pm$, respectively, $2\twida$, $2\twidb$ and
$\twidg$.\footnote {In $\bd(t) \to \pi^+\pi^-$ there may be
  significant penguin contributions \cite{penguins}, so that there are
  more than two interfering amplitudes. If this is the case, we assume
  that an isospin analysis is employed to find the CP-odd relative
  phase in the absence of penguins \cite{isospin}.} The trigonometric
functions of these phases which will be determined by the CP
asymmetries in these three decays will be $\sin 2\twida$, $\sin
2\twidb$ and $\sin^2 \twidg$ (or equivalently $\cos 2\twidg$),
respectively.

As the notation suggests, when new physics is absent, $\twida =
\alpha$, $\twidb = \beta$ and $\twidg = \gamma$. However, in the
presence of new physics, the situation changes. Most likely, if new
physics affects CP violation in the $B$ system, it does so by changing
the phase of the neutral $B$-${\bar B}$ mixing amplitudes \cite{DLN}.
In this paper, we shall assume that new physics enters only in this
way. In the presence of this new physics,
\beq
\label{thetad}
{\rm arg} \, A(B_q \to {\overline{B_q}}) = {\rm arg} \, A(B_q \to
{\overline{B_q}})|_{\sss SM} + 2 \theta_q ~;~~~~~ q=d,s ~.
\eeq
Here, $A(B_q \to {\overline{B_q}})$ is, of course, the $B_q \to
{\overline{B_q}}$ amplitude, ${\rm arg} \, A(B_q \to
{\overline{B_q}})|_{\sss SM} = 2 \, {\rm arg} \, (V_{tq} V_{tb}^*)$ is
the phase of $B_q \to {\overline{B_q}}$ mixing in the Standard Model,
and $2\theta_q$ is the change in this phase due to new physics. When
this new physics is present, the phases probed by $\bd(t) \to
\pi^+\pi^-$ and $\bd(t) \to \Psi \ks$ are changed to $\twida = \alpha
+ \theta_d$ and $\twidb = \beta - \theta_d$, respectively
\cite{nirsilv}. The phase $\twidg$ probed by $B^\pm \to D K^\pm$,
which does not involve neutral $B$ mixing, remains $\gamma$. The
situation is summarized in Table \ref{phasetable}.

\begin{table}
\hfil
\vbox{\offinterlineskip
\halign{&\vrule#&
\strut\quad#\hfil\quad\cr
\noalign{\hrule}
height2pt&\omit&&\omit&&\omit&&\omit&\cr
& Process && Relative Phase && CP Asymmetry && Value of & \cr
& \omit && of Amplitudes && Measures && Probed Phase& \cr
height2pt&\omit&&\omit&&\omit&&\omit&\cr
\noalign{\hrule}
height2pt&\omit&&\omit&&\omit&&\omit&\cr
& $\bd(t) \to \pi^+\pi^-$ && $2\twida$ && $\sin 2\twida$ &&
$\twida = \alpha + \theta_d$ & \cr
height2pt&\omit&&\omit&&\omit&&\omit&\cr
& $\bd(t) \to \Psi \ks$ && $2\twidb$ && $\sin 2\twidb$ &&
$\twidb = \beta - \theta_d$ & \cr
height2pt&\omit&&\omit&&\omit&&\omit&\cr
& $B^\pm \to D K^\pm$ && $\twidg$ && $\cos 2\twidg$ &&
$\twidg = \gamma$ & \cr
height2pt&\omit&&\omit&&\omit&&\omit&\cr
\noalign{\hrule}}}
\caption{The CP-violating phase information to be obtained from
first-round $B$ experiments on CP violation.}
\label{phasetable}
\end{table}

To test the SM, it is reasonable to assume, at least provisionally,
that no new physics is present, so that $(\twida,\twidb,\twidg) =
(\alpha,\beta,\gamma)$. Then $\twida$, $\twidb$ and $\twidg$ satisfy
the triangle conditions: $\twida$, $\twidb$ and $\twidg$ are all of
the same sign, and $|\twida+\twidb+\twidg| = \pi$. When this
assumption is made, the quantities $\sin 2\twida$, $\sin 2\twidb$ and
$\cos 2\twidg$ to be measured by the early $B$ CP experiments always
leave a twofold discrete ambiguity in the angle-set
$(\twida,\twidb,\twidg)$. As we shall see, at most one of these two
candidate solutions is consistent with the SM. In order to establish
the presence of new physics, it will be necessary to resolve the
ambiguity between the two solutions. In this paper, we will refer to
this as ``discrete ambiguity resolution'' (DAR).

Up to this point, the entire discussion has essentially been a review.
It has been known for quite some time now that the presence of
discrete ambiguities can make the discovery of new physics
problematic, and various aspects of issues related to this problem
have been discussed in the literature
\cite{NirQuinn,DA,DAWolf,GNW,GQ}. In Ref.~\cite{DAWolf}, Wolfenstein
presented a specific example of how a nonzero value of the new-physics
phase $\theta_d$ may be difficult to detect due to discrete
ambiguities. Grossman, Nir and Worah have proposed a method to measure
the phase (and magnitude) of the new-physics contribution to $B$
mixing \cite{GNW}. However, their technique runs into serious
difficulties due to discrete ambiguities. And Grossman and Quinn have
discussed a variety of methods of removing discrete ambiguities
\cite{GQ}.

What is missing from this collective analysis, however, is a complete,
systematic study of how the presence of new physics affects the
resolution of the discrete ambiguities. For example, some of the DAR
methods presented in Ref.~\cite{GQ} may not work if new physics is
present. It is therefore important to re-examine the various DAR
techniques, assuming the presence of new physics, to see exactly what
information they give us. (After all, if one assumes that new physics
is not present, then no DAR is necessary -- one simply chooses the
solution which is consistent with all other measurements.)

Furthermore, for arbitrary values of $\theta_d$ it is not always
obvious just how DAR will work to reveal the presence of new physics.
For example, as mentioned above, the measurements of $\sin 2\twida$,
$\sin 2\twidb$ and $\cos 2\twidg$ leave a twofold discrete ambiguity
in the angle-set $(\twida,\twidb,\twidg)$. If there is no new physics
($\theta_d=0$), then one angle-set will form a triangle which lies
within the allowed region of Fig.~\ref{rhoeta1} (the SM solution),
while the other lies outside. DAR will then clearly choose the SM
solution. If $\theta_d$ is relatively small, then the two candidate
triangles will be close to the $\theta_d=0$ triangles, and DAR will
choose that triangle which is close to the SM solution. However,
suppose now that $\theta_d$ is large. In this case, one or both of
$\twida$ and $\twidb$ may change sign relative to the original
$\alpha$ and $\beta$ (see Table \ref{phasetable}), so that the
angle-set $(\twida,\twidb,\twidg)$ will not even form a triangle. Will
DAR still reveal the presence of new physics in such a situation? And
if it does, how are the angles of the triangle thus found related to
the (non-triangle) angle-set $(\twida,\twidb,\twidg)$? These types of
questions must be investigated in order to understand how the presence
of new physics may affect the resolution of discrete ambiguities.

Finally, one might wonder how ``likely'' it is that DAR will be
necessary in order to discover new physics. That is, for what values
of $(\alpha,\beta,\gamma)$ and $\theta_d$ will we find at least one
solution which lies in the allowed region of Fig.~\ref{rhoeta1}?

In this paper we examine all of these issues, as well as some others.
We begin in Sec.~2 with a review of the discrete ambiguities which are
present when the quantities $\sin 2\twida$, $\sin 2\twidb$ and $\cos
2\twidg$ are measured. (Note that other analyses \cite{DAWolf,GNW,GQ}
usually assume that only $\sin 2\twida$ and $\sin 2\twidb$ have been
measured. However, since the measurement of the CP asymmetry in
$\bd(t) \to \pi^+\pi^-$ is likely to be at least as difficult as that
in $B^\pm \to D K^\pm$, it seems more reasonable to assume that both
will be measured.) When $(\twida,\twidb,\twidg)$
satisfy the triangle conditions of Eqs.~(\ref{tricond1}) and
(\ref{tricond2}), we show that these measurements always leave a
twofold discrete ambiguity in the angle-set $(\twida,\twidb,\twidg)$.
We will also show that at most one of these two candidate solutions is
consistent with the SM.

We also briefly examine how the various DAR methods are affected by
the presence of new physics. These methods can be separated into two
categories: those that measure $\sin 2\phi$ or $\cos 2\phi$, where
$\phi$ is some combination of CP phases, and those that measure $\sin
\phi$. These latter methods are dependent on theoretical assumptions.
As we shall see, in the presence of new physics, those methods which
probe $\sin \phi$ may not correctly resolve the discrete ambiguity.
However, those which measure $\sin 2\phi$ or $\cos 2\phi$ are still
reliable.

In Sec.~3, we examine the question of how the presence of physics
beyond the SM affects the search for this same new physics via the
resolution of discrete ambiguities. Assuming that the new physics
affects $\twida$ and $\twidb$ as indicated in Table \ref{phasetable},
$\twida$, $\twidb$ and $\twidg$ still satisfy one of the triangle
conditions: $|\twida+\twidb+\twidg| = \pi$. Thus, the presence of the
new physics would not be revealed by looking for a violation of this
condition. However, when $\theta_d$ is present, $\twida$, $\twidb$
and $\twidg$ may not all be of like sign, in violation of the other
triangle condition, Eq.~(\ref{tricond1}). If one could resolve the
discrete ambiguities in $\twida$, $\twidb$ and $\twidg$ separately,
without assuming that the phases satisfy the triangle conditions, then
when $\twida$, $\twidb$ and $\twidg$ are not of like sign one would
immediately discover that new physics is present. Unfortunately, as
indicated above, this is not possible: to fully resolve the discrete
ambiguities in a CP phase $\phi$, one needs $\sin\phi$, and the
methods which probe $\sin \phi$ are not reliable in the presence of
new physics.

Suppose, then, that one simply assumes provisionally that $\twida$,
$\twidb$ and $\twidg$ do satisfy both triangle conditions, and in
particular are of like sign. Under what circumstances would new
physics still be uncovered? In Sec.~3 we shall see that whether or not
$\theta_d$ results in $\twida$, $\twidb$ and $\twidg$ not being of
like sign, if one assumes that they {\it are} of like sign, the
measured quantities (Table \ref{phasetable}, Column 3) always lead to
two candidate solutions for $(\twida,\twidb,\twidg)$. There are then
three possibilities:
\begin{enumerate}

\item Both of the candidate $(\twida,\twidb,\twidg)$ solutions
  obtained assuming $\twida$, $\twidb$ and $\twidg$ are of like sign
  are consistent with the allowed $(\rho,\eta)$ region in
  Fig.~\ref{rhoeta1}. As mentioned above, in Sec.~2, we will see that,
  in practice, this is impossible.

\item One of the candidate $(\twida,\twidb,\twidg)$ solutions is
  consistent with the allowed $(\rho,\eta)$ region, but the other is
  not.

\item Neither of the candidate $(\twida,\twidb,\twidg)$ solutions is
  consistent with the allowed $(\rho,\eta)$ region. In this case, it
  is clear that new physics is present.

\end{enumerate}

Of these three possibilities, case (2) is obviously the one which
causes problems. Even if physics beyond the SM is present, due to the
existence of the twofold discrete ambiguity, the measurements of $\sin
2\twida$, $\sin 2\twidb$ and $\cos 2\twidg$ alone will not
unequivocally reveal its presence. In order to know whether or not new
physics is present, it will be necessary to remove the discrete
ambiguity.

When the true $(\twida,\twidb$, and $\twidg)$ are of like sign, one of
the two candidates for $(\twida,\twidb,\twidg)$ is the true
$(\twida,\twidb,\twidg)$. In this situation, the discrete ambiguity
resolution (DAR) techniques to be reviewed in Sec.~2 will select from
among the two candidates the true one. When the true $(\twida,\twidb$,
and $\twidg)$ are not of like sign, neither of the candidates for
$(\twida,\twidb,\twidg)$ is the true $(\twida,\twidb,\twidg)$, and the
DAR techniques simply select one or the other of the incorrect
candidates. Suppose, now, that in the case (2) where one of the
candidates is consistent with the allowed $(\rho,\eta)$ region but the
other is not, DAR selects as the alleged true solution the one which
is inconsistent with the allowed region.  Then, regardless of the
signs of the true $(\twida, \twidb$, and $\twidg)$, one would know for
certain that new physics is present, and one would not have known this
without the DAR. However, it might also happen that the DAR selects
the candidate solution which is consistent with the allowed
$(\rho,\eta)$ region. Then it could be that no new physics is present
and this solution represents the true angles $\alpha$, $\beta$ and
$\gamma$ in the unitarity triangle. But it could also be that new
physics {\it is} present, but that the CP angles and $\theta_d$ are
such that it remains hidden. In Sec.~3 we will provide illustrative
examples of all of these situations.

The above analysis is done assuming that both $\BK$ and $\BBd$ are
positive. However, since these theoretical predictions have not been
confirmed experimentally, it is conceivable that this assumption is
incorrect. In Sec.~4 we examine the consequences for DAR and the
search for new physics assuming that the signs of $\BK$ and $\BBd$ are
unknown. As we will show, under this assumption, DAR can {\it not}
definitively establish the presence of new physics. We conclude in
Sec.~5.

\section{Discrete Ambiguities and Their Resolution}

The CP asymmetries in the decays $\bd(t)\to\pi^+\pi^-$,
$\bd(t)\to\Psi\ks$ and $B^\pm \to D K^\pm$ permit the extraction of
the functions $\sin 2\twida$, $\sin 2\twidb$ and $\sin^2 \twidg$ (or
equivalently $\cos 2\twidg$), respectively\footnote{When an isospin
  analysis is used to extract $\sin 2\twida$ from the decay
  $\bd(t)\to\pi^+\pi^-$ despite the presence of penguins, the net
  result is that $\sin 2\twida$ is itself obtained with a fourfold
  discrete ambiguity, which depends on the relative magnitude and
  phase of the penguin and tree amplitudes. In this paper we ignore
  this ambiguity, since in general only one of the four values of
  $\sin 2\twida$ yields values of $\twida$ which can satisfy $|\twida
  + \twidb + \twidg| = \pi$. Furthermore, $\sin 2\twida$ can be
  extracted independently with no discrete ambiguity from a study of
  $B \to \rho\pi$ decays \cite{Dalitz}.}. (An alternative way of
getting at $\twidg$ is through the CP asymmetry in $\bs(t)\to D_s^\pm
K^\mp$ \cite{ADK}. However, even in this case, the function measured
is $\sin^2 \twidg$.)
%We will discuss this decay in more detail below.)
Thus, from these measurements, each CP angle can be obtained up to a
fourfold ambiguity: if $\twida_0$, $\twidb_0$ and $\twidg_0$ are the
true values of these angles, the values consistent with the
measurements are:
\begin{eqnarray}
&~& \twida_0 ~,~ \twida_0 + \pi ~,~ {\pi \over 2} - \twida_0 ~,~ -{\pi
\over 2} - \twida_0 ~, \nonumber \\
&~& \twidb_0 ~,~ \twidb_0 + \pi ~,~ {\pi \over 2} - \twidb_0 ~,~ -{\pi
\over 2} - \twidb_0 ~, \nonumber \\
&~& \twidg_0 ~,~ \twidg_0 + \pi ~,~ - \twidg_0 ~,~ - \twidg_0 - \pi ~.
\end{eqnarray}

There is thus a 64-fold discrete ambiguity in the extraction of the
CP-angle set $(\twida,\twidb,\twidg)$. However, assuming that the
three angles are the interior angles of a triangle, i.e.\ that they
satisfy Eqs.~(\ref{tricond1}) and (\ref{tricond2}), this discrete
ambiguity can be reduced to a twofold one. This result can be
found in the literature \cite{GQ}, although an explicit proof has not
been given. For completeness, we present such a proof in this paper.
However, since the main purpose of the paper is to examine the
consequences of this discrete ambiguity, we defer the proof of its
existence to Appendix A, and simply list the possible discrete
ambiguities in Table \ref{disambtable}.

\begin{table}
\hfil
\vbox{\offinterlineskip
\halign{&\vrule#&
\strut\quad#\hfil\quad\cr
\noalign{\hrule}
height2pt&\omit&&\omit&&\omit&\cr
& $Sign(\sin 2\twida)$ && $Sign(\sin 2\twidb)$ && Discrete Ambiguity & \cr
height2pt&\omit&&\omit&&\omit&\cr
\noalign{\hrule}
height2pt&\omit&&\omit&&\omit&\cr
& $>0$ && $>0$ && $\left(\twida, \twidb, \twidg \right) \to
\left({\pi \over 2} - \twida, {\pi \over 2} - \twidb, \pi - \twidg \right)$ 
& \cr
& $>0$ && $<0$ && $\left(\twida, \twidb, \twidg \right) \to
\left(-{\pi \over 2} - \twida, {\pi \over 2} - \twidb, - \twidg \right)$ & 
\cr
& $<0$ && $>0$ && $\left(\twida, \twidb, \twidg \right) \to
\left({\pi \over 2} - \twida, -{\pi \over 2} - \twidb, - \twidg \right)$ & 
\cr
& $<0$ && $<0$ && $\left(\twida, \twidb, \twidg \right) \to
\left(-{\pi \over 2} - \twida, -{\pi \over 2} - \twidb, -\pi - \twidg 
\right)$ & \cr
height2pt&\omit&&\omit&&\omit&\cr
\noalign{\hrule}}}
\caption{The twofold discrete ambiguity in $(\twida,\twidb,\twidg)$ 
remaining
after measurement of $\sin 2\twida$, $\sin 2\twidb$ and $\cos 2\twidg$.}
\label{disambtable}
\end{table}

There is one point which should be noted here. Within the SM, the
magnitude of the angle $\beta$ is constrained to be $16^\circ \le
|\beta| \le 35^\circ$ [Eq.~(\ref{betaconstraint})]. However, an
examination of Table \ref{disambtable} reveals that, regardless of
$Sign(\sin 2\twidb)$, at most one of the two $\twidb$ solutions
satisfies this constraint (and it can be that neither does).
Therefore, regardless of the signs of the candidate angle sets or of
$\BK$ and $\BBd$, {\it at most one of the two discretely ambiguous
  solutions can be consistent with the SM}. This will have important
consequences in our discussion of new physics.

If one assumes that there is no new physics (i.e.\ $\twida = \alpha$,
$\twidb = \beta$, $\twidg = \gamma$), this twofold discrete ambiguity
does not pose a problem. Since only one of the two solutions can be
consistent with the SM, then clearly that is the one which must be
chosen.

On the other hand, if one allows for the possibility that new physics
may affect the CP asymmetries, then there may be a problem. If both
solutions are inconsistent with the SM, then it is clear that new
physics is present. However, if one solution is consistent with the
SM, while the other is not, then the measurements of $\sin 2\twida$,
$\sin 2\twidb$ and $\cos 2\twidg$ alone will not tell us which of the
two solutions is the correct one. In other words, at this stage we
will not know whether or not new physics is present. In order to
decide, it will be important to be able to remove the discrete
ambiguities. Below, we briefly review various ways of doing this.

\subsection{Discrete Ambiguity Resolution (1)}

The methods for discrete ambiguity resolution can be separated into
two categories. First, techniques involving indirect, mixing-induced
CP violation can be used to extract trigonometric functions of the CP
phases other than $\sin 2\twida$, $\sin 2\twidb$ and $\cos 2\twidg$.
For example, $\cos 2\twida$ can be obtained through a study of the
time-dependent Dalitz plot for $\bd(t)\to\rho\pi$ decays
\cite{Dalitz}. Similarly, Dalitz-plot analyses of the decays $\bd(t)
\to D^+ D^- \ks$ and $\bd(t) \to D^\pm \pi^\mp \ks$ allow one to
extract the functions $\cos 2\twidb$ and $\sin 2(2\twidb + \twidg)$,
respectively \cite{Charlesetal}. $\cos 2\twidb$ can also be obtained
through a study of $\bd \to \Psi + K \to \Psi + (\pi^- \ell^+ \nu)$,
known as ``cascade mixing'' \cite{Azimov,cascade}. Finally, $\sin
2\twidg$ can be obtained from $\bs(t)\to D_s^\pm K^\mp$ if the width
difference between the two $B_s$ mass eigenstates is measurable
\cite{IsiBs}. For a more complete discussion of these techniques, we
refer to the reader to Refs.~\cite{GQ} and \cite{CPbook}.

Knowledge of any of these functions --- $\cos 2\twida$, $\cos 2\twidb$,
$\sin 2(2\twidb + \twidg)$, or $\sin 2\twidg$ --- is sufficient to
remove the discrete ambiguity left by the measurements of $\sin
2\twida$, $\sin 2\twidb$ and $\cos 2\twidg$. (There are also several
CP asymmetries that can be used to extract $\sin^2(2\twidb + \twidg)$:
$\bd(t) \to D^{(**)} \ks$, $\bd(t) \to D^\mp \pi^\pm$
\cite{neutralBtoDK}. However, it is straightforward to show that the
discrete ambiguities in Table \ref{disambtable} are {\it not} resolved
by this measurement.) In fact, if $\sin 2\twidg$ is known, the
measurement of $\cos 2\twidg$ is not even necessary. Assuming that
$\twida$, $\twidb$ and $\twidg$ obey the triangle conditions
[Eqs.~(\ref{tricond1}) and (\ref{tricond2})], except for certain
singular values of the CP angles, the measurements of $\sin 2\twida$,
$\sin 2\twidb$ and $\sin 2\twidg$ determine $(\twida,\twidb,\twidg)$
uniquely. This fact was also pointed out in Ref.~\cite{NirQuinn},
though the proof was not given. We present the proof in Appendix B.

Actually, the above analysis is not complete. If there is no new
physics, i.e.\ $\theta_d = 0$, then $\twida = \alpha$, $\twidb =
\beta$, and $\twidg = \gamma$. In this case it is true that, as
described above, each of the various methods does probe the given
trigonometric function of the CP angles, and the discrete ambiguity is
indeed resolved by any of the above measurements. However, as we have
argued earlier, this is not particularly relevant -- if no new physics
is assumed, there is no need for DAR. What we really want to know is
the following: in the presence of new physics, do the above methods
still probe the same functions of the CP angles as they did in the
absence of new physics? That is, are we certain that we have taken
into account all possible large new-physics effects?

In fact, for this class of DAR methods, the answer to these questions
is yes. Even in the presence of new physics, each technique probes the
same function of CP angles as it did in the absence of new physics,
with (obviously) $\alpha$ being replaced by $\twida$, and similarly
for $\beta$ and $\gamma$. We demonstrate this explicitly for cascade
mixing.

The decay $\bd \to \Psi + (\pi^- \ell^+ \nu)_K$ can occur via the path
$\bd \to \Psi + K^0 \to \Psi + (\pi^- \ell^+ \nu)_K$ and the path $\bd
\to \bdbar \to \Psi + \overline{K^0} \to \Psi + K^0 \to \Psi + (\pi^-
\ell^+ \nu)_K$. In the latter path, neutral $K$ mixing follows neutral
$B$ mixing. In the absence of new physics, the interference between
the two paths probes $\sin 2\beta$ and $\cos 2\beta$.  Now, new
physics can modify both the phase and the magnitude of the $\bd \to
\bdbar$ mixing amplitude. Since, in the presence of new physics, $\arg
A(B_d \to \overline{B_d}) = -2 (\beta - \theta_d) = -2\twidb$, the
first of these effects is taken into account by replacing $\beta$ by
$\twidb$ in the expression for the decay rate. The second is taken
into account by using the physical neutral $B$ mass splitting. New
physics can conceivably also affect $K^0 \to \overline{K^0}$ mixing,
but this is taken into account by using the physical neutral $K$ mass
splitting and decay widths in the expression for the rate.  The
CP-violating phase of $K^0 \to \overline{K^0}$ mixing is known to be
small, whether it involves new physics or not. In particular, it is
negligible compared to the effects expected in the $B$ system.
Finally, significant new-physics effects on the $B \to \Psi K$ decay
amplitudes are not expected. Thus, the replacement of $\beta$ by
$\twidb$ and the use of the observed mass splittings and widths takes
into account all significant new-physics effects.

Although we will not show it explicitly, it is not difficult to
convince oneself that a similar logic holds for the other DAR methods
which rely on indirect CP violation. In all cases, the only place
where there can be significant new-physics effects is in $B$ mixing,
and this is taken into account by the replacement of $\alpha$, $\beta$
and $\gamma$ by $\twida$, $\twidb$ and $\twidg$, respectively. We
therefore conclude that these methods can be reliably used to remove
the discrete ambiguity even in the presence of new physics.

In the illustrative case of cascade mixing, one finds by explicit 
calculation that
\cite{cascade}
\begin{eqnarray}
&~& \Gamma \left[ \bd \to \Psi + K \to \Psi + (\pi^- \ell^+ \nu) \right]
\nonumber \\
&~& \qquad\qquad
\propto e^{-\Gamma_{\sss B} \tau_{\sss B}} \left\{
e^{-\gamma_{\sss S}\tau_{\sss K}} \left[ 1 - \sin 2\twidb \, \sin(\Delta
M_{\sss B} \tau_{\sss B}) \right] \right. \nonumber \\
&~& \qquad\qquad\qquad ~~~~~
+ e^{-\gamma_{\sss L}\tau_{\sss K}} \left[ 1 + \sin 2\twidb \, \sin(\Delta
M_{\sss B} \tau_{\sss B}) \right] \nonumber \\
&~& \qquad\qquad\qquad
+ \, 2 \, e^{-{1\over 2} ( \gamma_{\sss S} + \gamma_{\sss L} )
\tau_{\sss K}} \left[ \cos (\Delta M_{\sss B} \tau_{\sss B})
\, \cos (\Delta M_{\sss K} \tau_{\sss K}) \right. \nonumber \\
&~& \qquad\qquad\qquad ~~~~~ \left.
+ \cos 2\twidb \,
\sin (\Delta M_{\sss B} \tau_{\sss B}) \,
\sin (\Delta M_{\sss K} \tau_{\sss K}) \right] \Bigr\} ~.
\label{eq12}
\end{eqnarray}
Here, $\Gamma_B$ is the width of the $B$, $\gamma_{\sss S}$ and
$\gamma_{\sss L}$ are, respectively, the widths of $\ks$ and $\kl$,
$\Delta M_{\sss B}$ and $\Delta M_{\sss K}$ are the positive $B$ and
$K$ mass splittings, and $\tau_{\sss B}$ and $\tau_{\sss K}$ are the
proper times that the $B$ and $K$ live before decay. A similar
expression holds when the initial state is a $\bdbar$, or the final
state is $\Psi + (\pi^+ \ell^- {\bar\nu})$.

Note that, although the expression of the rate for $\bd \to \Psi +
(\pi^- \ell^+ \nu)$ is more complicated than that for $\bd \to \Psi
\ks$, it is still independent of hadronic uncertainties. It depends
only on the CP angle $\twidb$, and on the known quantities
$\Gamma_{\sss B}$, $\gamma_{\sss S}$, $\gamma_{\sss L}$, $\Delta
M_{\sss B}$ and $\Delta M_{\sss K}$. The key point is that the
function $\cos 2\twidb$ appears in the expression for the rate. Thus,
this method allows one to measure $\cos 2\twidb$ and thus remove the
discrete ambiguity of Table \ref{disambtable}.

\subsection{Discrete Ambiguity Resolution (2)}

The second category of DAR methods are those which use direct CP
violation, and probe the sine of some CP phase \cite{GQ}. For example,
recall that CP violation in $\bd(t) \to \Psi\ks$ probes $\sin
2\twidb$. In principle, the decay $\bd(t) \to D^+ D^-$ can also be
used to extract $\sin 2\twidb$. However, this decay may not be
dominated by a single decay amplitude --- there may be a significant
penguin contribution --- so the CP-violation measurement will not be
free of hadronic uncertainties. Still, by comparing the two
measurements and using some theoretical input, one can obtain $\cos
2\twidb \sin\twidb$, or at least its sign. Similarly a comparison of
CP violation in $\bd(t) \to \rho\pi$ and $\bd(t) \to \pi^+\pi^-0$
gives information about $\cos 2\twida \sin \twida$. Any such
information may help in reducing the discrete ambiguities.

Unfortunately, there are problems with such methods. Even leaving
aside the question of the theoretical input, if there is new physics,
these methods cannot be reliably used for discrete ambiguity
resolution. The reason is that new physics can affect not only $B$
mixing, but also the penguin contributions to $B$ decays
\cite{NPpenguins}. As before, the easiest way to see this is with an
example.

The decay $\bd \to D^+ D^-$ receives contributions from a tree
amplitude and a $b\to d$ penguin amplitude, which itself has
contributions from internal $u$, $c$ and $t$-quarks. Eliminating the
$u$-quark penguin piece via CKM unitarity, and combining the $c$-quark
contribution with the tree contribution, in the Wolfenstein
approximation the total amplitude for $\bd \to D^+ D^-$ can be
written
\begin{eqnarray}
A & = & A_{\sss T} e^{i \delta_T}
  + A_{\sss P} e^{i \delta_P} e^{-i \beta'} ~, \nonumber \\
{\bar A} & = & A_{\sss T} e^{i \delta_T}
  + A_{\sss P} e^{i \delta_P} e^{i \beta'} ~.
\end{eqnarray}
Here we have written the weak phase of the $t$-quark penguin
contribution as $\beta'$. In the SM, this phase is $\beta$. However,
it is conceivable that new physics can contribute to the $b\to d$
penguin, thereby altering this phase.

By measuring the time-dependent rate for $\bd(t) \to D^+ D^-$, one can
extract the CP asymmetry $a_{D^+D^-}^{\rm sin}$, defined as
\beq
a_{D^+D^-}^{\rm sin} \equiv
{-2 \, {\rm Im} \lambda \over 1 + |\lambda|^2} ~,
\eeq
where
\beq
\lambda \equiv {q\over p} \, {{\bar A} \over A} ~.
\eeq
Here $q/p$ is the phase of $B$ mixing. Allowing for the presence of
new physics in the mixing, this can be written as
\beq
{q\over p} = e^{-2 i \twidb} ~.
\eeq

For the final state $D^+ D^-$, the CP asymmetry is
\beq
a_{D^+D^-}^{\rm sin} \simeq \sin 2\twidb
    - 2 r \cos\delta \cos 2\twidb \sin \beta' ~,
\eeq
where $\delta \equiv \delta_{\sss P} - \delta_{\sss T}$, $r\equiv
A_{\sss P}/A_{\sss T}$, and we have neglected terms of $O(r^2)$. The
function $\sin 2\twidb$ is assumed to have been measured in $\bd(t)
\to \Psi \ks$ and it is also assumed that we have some reliable
theoretical input concerning the hadronic factor $r\cos\delta$. Now,
in the absence of new physics, $\twidb = \beta' = \beta$. Thus, the
measurement of $a_{D^+D^-}^{\rm sin}$ would allow us to obtain
information about $\cos 2\beta \sin \beta$. However, this no longer
holds in the presence of new physics. Any new physics which
contributes to $\bd$-$\bdbar$ mixing will, in general, also contribute
to the $b\to d$ penguin. However, it will not necessarily contribute
in the same way. That is, there is no reason to expect that $\twidb =
\beta'$. In particular, it is conceivable that the new physics is such
that $\twidb$ and $\beta'$ are of opposite sign, in which case even
knowledge of the sign of $\cos 2\twidb \sin\beta'$ may not tell us
anything about the sign of $\twidb$. We therefore conclude that this
method, and others like it, cannot be reliably used to remove discrete
ambiguities if there is new physics. (Similar conclusions have been
reached in Refs.~\cite{GQ} and \cite{DAWolf}.)

\subsection{Reduction of the Allowed $(\rho,\eta)$ Region}

Finally, we reiterate that the main problem caused by the presence of
the twofold discrete ambiguity is the possibility of having one
solution inside the allowed region of Fig.~\ref{rhoeta1}, and the
other outside. In this case, one does not know whether or not new
physics is present. Above we reviewed the measurements which can be
used to remove the discrete ambiguity. However, there is another
approach which can be used. If the allowed region of
Fig.~\ref{rhoeta1} were reduced, this would then reduce the likelihood
of having one solution inside the allowed region in the presence of
new physics.

The measurements which contribute to the allowed region of
Fig.~\ref{rhoeta1} are $|V_{cb}|$, $|V_{ub}/V_{cb}|$, $B_d$ and $B_s$
mixing, and CP violation in the kaon system ($\epsilon_K$). If the
error on any of these measurements can be reduced, either through
reduced experimental error or better understanding of the theoretical
uncertainties, this would help to reduce the allowed region. In fact,
over the past year or two, the improved lower bound on $B_s$ mixing
has already removed roughly half of the previously-allowed
$(\rho,\eta)$ region. An actual value for this mixing would be helpful
indeed.

There is another measurement which can help to reduce the allowed
$(\rho,\eta)$ region. Within the SM, the decay $\kl \to \pi^0 \nu
\bar\nu$ probes the Wolfenstein parameter $\eta$. In the Wolfenstein
parametrization, the branching ratio can be written \cite{BurFleisch}:
\beq
B(\kl \to \pi^0 \nu {\bar\nu} ) = 3.0 \times 10^{-11} \left[ {\eta \over
0.39} \right]^2 \left[ {{\bar m}_t(m_t) \over 170~{\rm GeV} }
\right]^{2.3}
\left[ { |V_{cb}| \over 0.040 } \right]^4 ~.
\eeq
Since the branching ratio is proportional to $\eta^2$, its measurement
would greatly help in constraining the allowed $(\rho,\eta)$ region.

Note that the precision with which $\eta$ can be extracted from a
measurement of $B(\kl \to \pi^0 \nu {\bar\nu} )$ is mainly limited by
the error on $|V_{cb}|$, which presently stands at about 4.3\%
\cite{PDG}, including both the experimental and theoretical (HQET)
errors. If the error on $|V_{cb}|$ could be reduced, then this
measurement could be used to obtain $|\eta|$, the height of the
unitarity triangle.

\section{Finding New Physics} \label{newphysics}

Let us now turn to the question of what specifically is learned when
discrete ambiguity resolution is applied. As usual, we assume that CP
violation in the $B$ system is affected by new physics (NP) beyond the
Standard Model. In particular, we assume that the NP affects
CP-violating asymmetries by modifying the phase of $\bd$-$\bdbar$
mixing, as described in the Introduction [Eq.~(\ref{thetad})]. The
phases probed by the first-round CP experiments on the $B$ system are
then the quantities $\twida=\alpha+\theta_d$, $\twidb=\beta-\theta_d$
and $\twidg=\gamma$ given in the last column of Table
\ref{phasetable}.

As we have noted, while $|\twida + \twidb + \twidg| = \pi$, in the
presence of a nonzero $\theta_d$, $\twida$, $\twidb$ and $\twidg$ may
not all be of the same sign, unlike the true angles $\alpha$, $\beta$
and $\gamma$ in the unitarity triangle. Resolving the discrete
ambiguities in $\twida$, $\twidb$ and $\twidg$ separately, and finding
that they are not all of like sign, would reveal the presence of NP.
However, as already mentioned, we do not have a method for reliably
resolving the ambiguities in $\twida$, $\twidb$ and $\twidg$
separately. Therefore, we ask under what conditions the NP would still
be visible even if one proceeds by simply {\it assuming} that
$\twida$, $\twidb$ and $\twidg$ satisfy both of the ``triangle
conditions'' [Eqs.~(\ref{tricond1}) and (\ref{tricond2})].

To answer this question, we first show that, given measured values of
$\sin 2\twida$, $\sin 2\twidb$ and $\cos 2\twidg$, there are always
two candidate solutions for $\twida$, $\twidb$ and $\twidg$ which
satisfy both triangle conditions. We call such solutions ``triangle
angle sets.'' To demonstrate that two of them always exist, we
distinguish two possibilities:
\begin{enumerate}

\item Suppose that, even with a nonzero $\theta_d$, $\twida$, $\twidb$
  and $\twidg$ are all of the same sign (that is, positive). Then the
  angle set ($\twida,\twidb,\twidg$) clearly satisfies the two
  triangle conditions of Eqs.~(\ref{tricond1}) and (\ref{tricond2}). There
  is then a second triangle angle set with the same values of the
  measured quantities. It is related to ($\twida,\twidb,\twidg$) by
  one of the entries in Table \ref{disambtable}. Note that one of
  these two triangle angle sets is obviously the true
  ($\twida,\twidb,\twidg$).

\item Suppose, instead, that $\theta_d$ is such that $\twida$,
  $\twidb$ and $\twidg$ are not of the same sign. This can occur in a
  number of ways. For example, $\theta_d$ can be such that $\twida =
  \alpha + \theta_d$ satisfies $-\pi < \twida < 0$, while $\twidb =
  \beta - \theta_d$ satisfies $0 < \twidb < \pi$. Or perhaps the
  result is $0 < \twida < \pi$ and $-\pi < \twidb < 0$. We refer to
  these situations, where $\theta_d$ has flipped the sign of $\alpha$
  or $\beta$, but not both, as a ``single flip.'' (Of course, $\twidg
  = \gamma$ is unaffected by $\theta_d$, and so remains positive.)  It
  is also possible that $\theta_d$ flips the sign of both $\alpha$ and
  $\beta$ -- a ``double flip.'' In all cases, one ends up with two of
  the three CP angles being of like sign, with the third having the
  opposite sign.

  In all of these cases, to obtain a triangle angle set, one has
  simply to add $\pm\pi$ to the two angles which have the same sign.
  This will give three same-sign angles which form a triangle and
  reproduce the measured quantities. And, as above, the second
  triangle angle set is obtained from the appropriate entry in Table
  \ref{disambtable}. However, in contrast to the previous case,
  neither of these two candidate triangle angle sets is the true
  ($\twida,\twidb,\twidg$).

\end{enumerate}
In Table \ref{fliptable} we summarize the relation between the true
$(\twida,\twidb,\twidg)$ and the triangle set
$(\twida_1,\twidb_1,\twidg_1)$ most simply related to it.

\begin{table}
\hfil
\vbox{\offinterlineskip
\halign{&\vrule#&
\strut\quad#\hfil\quad\cr
\noalign{\hrule}
height2pt&\omit&&\omit&\cr
& Angle(s) flipped && $(\twida_1,\twidb_1,\twidg_1$) & \cr
height2pt&\omit&&\omit&\cr
\noalign{\hrule}
height2pt&\omit&&\omit&\cr
& none && $\left(\twida, \twidb, \twidg \right)$ & \cr
& $\alpha$ && $\left(\twida, \twidb - \pi, \twidg - \pi \right)$ & \cr
& $\beta$ && $\left(\twida - \pi, \twidb, \twidg - \pi\right)$ & \cr
& $\alpha$, $\beta$ (1) && $\left(\twida - \pi, \twidb + \pi, \twidg
\right)$ & \cr
& $\alpha$, $\beta$ (2) && $\left(\twida + \pi, \twidb - \pi, \twidg
\right)$ & \cr
height2pt&\omit&&\omit&\cr
\noalign{\hrule}}}
\caption{Construction of the triangle angle set
$(\twida_1,\twidb_1,\twidg_1)$
  in the presence of NP. The second triangle angle set
  $(\twida_2,\twidb_2,\twidg_2)$ can be obtained from the appropriate
  entry in Table \protect\ref{disambtable}. (1) refers to the case
  where $\pi < \twida < 2 \pi$ and $-\pi < \twidb < 0$; (2) refers to
  $-\pi < \twida < 0$ and $\pi < \twidb < 2 \pi$.}
\label{fliptable}
\end{table}

In summary, if NP enters by modifying $\bd$-$\bdbar$ mixing, then
there are always two triangle angle sets,
$(\twida_1,\twidb_1,\twidg_1)$ and $(\twida_2,\twidb_2,\twidg_2)$,
which give {\it any} observed values of the measured quantities. These
two sets are related by one of the entries in Table \ref{disambtable},
depending on the signs of $\sin 2\twida$ and $\sin 2\twidb$. If the
true $\twida$, $\twidb$ and $\twidg$ are of like sign (no flips), then
one of these two angle sets, which we call
$(\twida_1,\twidb_1,\twidg_1)$, is the true $(\twida,\twidb,\twidg)$.
If the true $\twida$, $\twidb$ and $\twidg$ are not of like sign (one
or two flips), then obviously neither $(\twida_1,\twidb_1,\twidg_1)$
nor $(\twida_2,\twidb_2,\twidg_2)$ is the true
$(\twida,\twidb,\twidg)$.

Suppose, now, that $\sin 2\twida$, $\sin 2\twidb$ and $\cos 2\twidg$
are measured. Suppose further that one then proceeds by assuming the
underlying angles $\twida$, $\twidb$ and $\twidg$ to be the angles
$\alpha$, $\beta$ and $\gamma$ in the unitarity triangle, and looking
for inconsistencies. One is then assuming that
$(\twida,\twidb,\twidg)$ is a triangle angle set, so one would
calculate that $(\twida,\twidb,\twidg)$ is either the set of angles we
have called $(\twida_1,\twidb_1,\twidg_1)$, or the one we have called
$(\twida_2,\twidb_2,\twidg_2)$. As already stated in the Introduction,
one will then encounter one of the following three situations:
\begin{enumerate}

\item Both $(\twida_1,\twidb_1,\twidg_1)$ and
  $(\twida_2,\twidb_2,\twidg_2)$ are consistent with the allowed
  $(\rho,\eta)$ region in Fig.~\ref{rhoeta1}.

\item Only one of $(\twida_1,\twidb_1,\twidg_1)$ and
  $(\twida_2,\twidb_2,\twidg_2)$ is consistent with the allowed
  $(\rho,\eta)$ region.

\item Neither $(\twida_1,\twidb_1,\twidg_1)$ nor
  $(\twida_2,\twidb_2,\twidg_2)$ is consistent with the allowed
  $(\rho,\eta)$ region.

\end{enumerate}
As we have already argued in Sec.~2, in practice situation (1) can
never occur. Only one of the two solutions related as in
Table~\ref{disambtable} will be consistent with known physics.
However, if new physics is indeed present, then resolution of the
discrete ambiguity will in fact establish its presence, except in some
specific cases which we describe below.

We are assuming that new physics affects CP violation in the $B$
system only by changing the phase of neutral $B$ mixing. We have seen
that when this is the case, there are always two candidate triangle
angle sets consistent with any given values of the measured quantities
$\sin 2\twida$, $\sin 2\twidb$ and $\cos 2\twidg$. Thus, if no other
quantities are measured, one cannot uncover the presence of the NP by
trying to determine whether the angles underlying the measured $\sin
2\twida$, $\sin 2\twidb$ and $\cos 2\twidg$ form a triangle angle set.
Can one, alternatively, uncover it by determining that the candidate
triangle angle sets correspond to points $(\rho,\eta)$ which are
outside the allowed region? A study of Table \ref{fliptable} shows
that the answer to this question is usually yes.

First, we note that, in Table \ref{fliptable}, $\twida_1$ is always
equal to $\twida$, $\twida + \pi$ or $\twida - \pi$, and similarly for
$\twidb_1$ and $\twidg_1$. The second angle set
$(\twida_2,\twidb_2,\twidg_2)$ is obtained from the appropriate entry
in Table \ref{disambtable}. This is important when one considers the
resolution of the twofold discrete ambiguity. In Sec.\ 2, we reviewed
various measurements which can be reliably used for DAR. In all cases,
one of the following trigonometric functions is obtained: $\cos
2\twida$, $\cos 2\twidb$, $\sin 2\twidg$ or $\sin 2(2\twidb +
\twidg)$. The key point is that all of these functions are unchanged
when one replaces $(\twida,\twidb,\twidg)$ by
$(\twida_1,\twidb_1,\twidg_1)$. On the other hand, these functions
change sign if $(\twida_2,\twidb_2,\twidg_2)$ is used. In other words,
{\it in all cases, the resolution of the discrete ambiguity chooses
  the angle set $(\twida_1,\twidb_1,\twidg_1)$}.

From this we conclude that discrete ambiguity resolution will always
reveal the presence of new physics unless
$(\twida_1,\twidb_1,\twidg_1)$ corresponds to a triangle which falls
within the allowed region of Fig.~\ref{rhoeta1}. And, by studying
Table \ref{fliptable}, it is straightforward to see when this occurs.
Whenever there is one flip (i.e.\ the second and third lines of Table
\ref{fliptable}), one ends up with a downward-pointing unitarity
triangle, which is always a signal of new physics (recall that we are
assuming that $\BK$ is positive). If there are no flips (i.e.\ the
first line of Table \ref{fliptable}), the resultant triangle will lie
within the allowed region if $(\twida,\twidb,\twidg) \simeq
(\alpha,\beta,\gamma)$. This can only happen if $\theta_d$ is small.
Similarly, if there are two flips (i.e.\ the fourth and fifth lines of
Table \ref{fliptable}), the resultant triangle will be consistent with
other measurements only if $\theta_d \pm \pi$ is small. We therefore
conclude that DAR will always reveal the presence of new physics
unless the new-physics angle $\theta_d$ is near 0 or $\pi$. (And note:
since only the quantity $2\theta_d$ ever enters into CP asymmetries
[see Eq.~(\ref{thetad})], a new-physics phase $\theta_d \simeq \pi$ is
essentially equivalent to $\theta_d \simeq 0$. In other words, in both
cases, we are effectively talking about small new-physics effects.)

Of course, this raises the following question: what precisely does
``near 0'' mean? The easiest way to answer this is to refer to
Eq.~\ref{betac2}, which gives the currently-allowed range for $\beta$
as $16^\circ \le \beta \le 35^\circ$. Depending on the true value of
$\beta$, for the case of $\theta_d \simeq 0$, $\theta_d$ must be small
enough that the value of $\twidb = \beta - \theta_d$ still lies within
this range. In such a situation the new-physics triangle will still
lie within the allowed region of Fig.~\ref{rhoeta1} and DAR will not
discover the presence of new physics. Similarly, if $\theta_d \simeq
\pi$, $\twidb$ must lie within the range $-164^\circ \le \twidb \le
-145^\circ$. In this way $\twidb + \pi$ will fall in the allowed range
for $\beta$, and again DAR will not reveal that new physics is
present. This shows explicitly that a reduction of the allowed
$(\rho,\eta)$ region, which would reduce the allowed range for
$\beta$, would help considerably in the search for new physics.

Once the discrete ambiguities have been removed, the method of
Grossman, Nir and Worah (GNW) \cite{GNW} can in principle be used to
measure the new-physics phase $\theta_d$. Among other things, this
method assumes that we know the new-physics triangle,
defined by the angles $(\twida,\twidb,\twidg)$. However, as we have
noted above, even when DAR does reveal the presence of new physics, it
does not always choose the true values of $(\twida,\twidb,\twidg)$. In
such cases, the GNW method will not find the true value of $\theta_d$,
but rather $\theta_d + \pi$. Still, as noted above (and in
Ref.~\cite{GNW}), as far as $B$ mixing is concerned, this difference
is not physical.

Above we have seen that there is uncertainty about the presence of new
physics only if one of the two discretely ambiguous solutions is
consistent with the allowed $(\rho,\eta)$ region, while the other is
not. And we have seen that DAR is needed to remove this uncertainty.
But this also suggests that, for other values of $\theta_d$, both
solutions will lie outside the allowed $(\rho,\eta)$ region. In such
cases, we will know that new physics is present, even without DAR. The
obvious question then is: how likely is it that DAR will be needed?

While we cannot answer this question in any statistically rigorous
way, one can still get a sense of how things work by performing a scan
of the $(\alpha,\beta,\gamma)$ and $\theta_d$ parameter space. The
results of such an analysis are shown in Tables \ref{E1}, \ref{E2},
\ref{E3}, \ref{E4} and \ref{E5}, which correspond, respectively, to
the assumption that the true angles in the unitarity triangle
correspond to a point $(\rho,\eta)$ which is near the center of the
$(\rho,\eta)$ region allowed by Fig.~\ref{rhoeta1}, near the left edge
of this region, right edge, the top, and the bottom\footnote{In all of
  our examples, we assume that the true $(\alpha,\beta,\gamma)$
  correspond to a $(\rho, \eta)$ within or very near the allowed
  $(\rho,\eta)$ region of Fig.~\ref{rhoeta1}. However, it should be
  noted that since the new physics affects $B$ mixing, which is one of
  the inputs to Fig.~\ref{rhoeta1}, that region may not represent the
  true allowed $(\rho,\eta)$ region. In other words, in the presence
  of new physics, the true SM $(\alpha,\beta,\gamma)$ may already lie
  outside the region of Fig.~\ref{rhoeta1}. In this paper, we do not
  consider this additional possibility, but its inclusion would not
  affect our conclusions.}. For each of these five sample cases, we
explore the effect of a $\theta_d$ of $22^\circ$, $45^\circ$,
$78^\circ$ and $90^\circ$. We give for each $\theta_d$ the two
candidate triangle angle sets, $(\twida_1,\twidb_1,\twidg_1)$ and
$(\twida_2,\twidb_2,\twidg_2)$, that would be inferred from the
measured $\sin 2\twida$, $\sin 2\twidb$ and $\cos 2\twidg$. Next to
each angle set, we give the corresponding inferred $(\rho,\eta)$
point.

\begin{table}
\vbox{\tabskip=0pt \offinterlineskip
\def\tablerule{\noalign{\hrule}}
\halign to500pt{\strut#&
\vrule#\tabskip=1em plus 2em&
\hfil#& \vrule\,\vrule#&
\hfil#\hfil& \vrule#&
\hfil#\hfil& \vrule\,\vrule#&
\hfil#\hfil& \vrule#&
\hfil#& \vrule#\tabskip=0pt\cr\tablerule
\omit&height2pt&\omit&&\omit&&\omit&&\omit&&\omit&\cr
&&\omit\hidewidth $\theta_d$ \hidewidth&&
\omit\hidewidth $\twida_1$, $\twidb_1$, $\twidg_1$ \hidewidth&&
\omit\hidewidth $(\rho,\eta)$ \hidewidth&&
\omit\hidewidth $\twida_2$, $\twidb_2$, $\twidg_2$ \hidewidth&&
\omit\hidewidth $(\rho,\eta)$ \hidewidth&\cr
\omit&height2pt&\omit&&\omit&&\omit&&\omit&&\omit&\cr
\tablerule
\omit&height2pt&\omit&&\omit&&\omit&&\omit&&\omit&\cr
&& $22^\circ$ && $117^\circ$, $3^\circ$, $60^\circ$ && $(0.03,0.05)$
&& $-27^\circ$, $-93^\circ$, $-60^\circ$ && $(1.10,-1.90)$ & \cr
&& $45^\circ$ && $-40^\circ$, $-20^\circ$, $-120^\circ$ && $(-0.27,-0.46)$
&& $-50^\circ$, $-70^\circ$, $-60^\circ$ && $(0.61,-1.06)$ & \cr
&& $68^\circ$ && $-17^\circ$, $-43^\circ$, $-120^\circ$ && $(-1.17,-2.02)$
&& $-73^\circ$, $-47^\circ$, $-60^\circ$ && $(0.38,-0.66)$ & \cr
&& $90^\circ$ && $5^\circ$, $115^\circ$, $60^\circ$ && $(5.20,9.01)$
&& $-95^\circ$, $-25^\circ$, $-60^\circ$ && $(0.21,-0.37)$ & \cr
\omit&height2pt&\omit&&\omit&&\omit&&\omit&&\omit&\cr
\tablerule }}
\caption{The effects of new physics when the true angles in the
unitarity triangle are $(\alpha,\beta,\gamma) =
(95^\circ,25^\circ,60^\circ)$, corresponding to
$(\rho,\eta)=(0.21,0.37)$.}
\label{E1}
\end{table}

\begin{table}
\vbox{\tabskip=0pt \offinterlineskip
\def\tablerule{\noalign{\hrule}}
\halign to500pt{\strut#&
\vrule#\tabskip=1em plus 2em&
\hfil#& \vrule\,\vrule#&
\hfil#\hfil& \vrule#&
\hfil#\hfil& \vrule\,\vrule#&
\hfil#\hfil& \vrule#&
\hfil#& \vrule#\tabskip=0pt\cr\tablerule
\omit&height2pt&\omit&&\omit&&\omit&&\omit&&\omit&\cr
&&\omit\hidewidth $\theta_d$ \hidewidth&&
\omit\hidewidth $\twida_1$, $\twidb_1$, $\twidg_1$ \hidewidth&&
\omit\hidewidth  $(\rho,\eta)$ \hidewidth&&
\omit\hidewidth  $\twida_2$, $\twidb_2$, $\twidg_2$ \hidewidth&&
\omit\hidewidth $(\rho,\eta)$ \hidewidth&\cr
\omit&height2pt&\omit&&\omit&&\omit&&\omit&&\omit&\cr
\tablerule
\omit&height2pt&\omit&&\omit&&\omit&&\omit&&\omit&\cr
&& $22^\circ$ && $-83^\circ$, $-2^\circ$, $-95^\circ$ && $(0.0,-0.04)$
&& $-7^\circ$, $-88^\circ$, $-85^\circ$ && $(0.71,-8.17)$ & \cr
&& $45^\circ$ && $-60^\circ$, $-25^\circ$, $-95^\circ$ && $(-0.04,-0.49)$
&& $-30^\circ$, $-65^\circ$, $-85^\circ$ && $(0.16,-1.81)$ & \cr
&& $68^\circ$ && $-37^\circ$, $-48^\circ$, $-95^\circ$ && $(-0.11,-1.23)$
&& $-53^\circ$, $-42^\circ$, $-85^\circ$ && $(0.07,-0.83)$ & \cr
&& $90^\circ$ && $-15^\circ$, $-70^\circ$, $-95^\circ$ && $(-0.32,-3.62)$
&& $-75^\circ$, $-20^\circ$, $-85^\circ$ && $(0.03,-0.35)$ & \cr
\omit&height2pt&\omit&&\omit&&\omit&&\omit&&\omit&\cr
\tablerule }}
\caption{Same as Table \protect\ref{E1}, but with the true
$(\alpha,\beta,\gamma) = (75^\circ,20^\circ,85^\circ)$, corresponding to
$(\rho,\eta)=(0.03,0.35)$.}
\label{E2}
\end{table}

\begin{table}
\vbox{\tabskip=0pt \offinterlineskip
\def\tablerule{\noalign{\hrule}}
\halign to510pt{\strut#&
\vrule#\tabskip=1em plus 2em&
\hfil#& \vrule\,\vrule#&
\hfil#\hfil& \vrule#&
\hfil#\hfil& \vrule\,\vrule#&
\hfil#\hfil& \vrule#&
\hfil#& \vrule#\tabskip=0pt\cr\tablerule
\omit&height2pt&\omit&&\omit&&\omit&&\omit&&\omit&\cr
&&\omit\hidewidth $\theta_d$ \hidewidth&&
\omit\hidewidth $\twida_1$, $\twidb_1$, $\twidg_1$ \hidewidth&&
\omit\hidewidth  $(\rho,\eta)$ \hidewidth&&
\omit\hidewidth  $\twida_2$, $\twidb_2$, $\twidg_2$ \hidewidth&&
\omit\hidewidth $(\rho,\eta)$ \hidewidth&\cr
\omit&height2pt&\omit&&\omit&&\omit&&\omit&&\omit&\cr
\tablerule
\omit&height2pt&\omit&&\omit&&\omit&&\omit&&\omit&\cr
&& $22^\circ$ && $132^\circ$, $8^\circ$, $40^\circ$ && $(0.14,0.12)$
&& $-42^\circ$, $-98^\circ$, $-40^\circ$ && $(1.13,-0.95)$ & \cr
&& $45^\circ$ && $-25^\circ$, $-15^\circ$, $-140^\circ$ && $(-0.47,-0.39)$
&& $-65^\circ$, $-75^\circ$, $-40^\circ$ && $(0.82,-0.69)$ & \cr
&& $68^\circ$ && $-2^\circ$, $-38^\circ$, $-140^\circ$ &&
$(-13.51,-11.34)$
&& $-88^\circ$, $-52^\circ$, $-40^\circ$ && $(0.60,-0.51)$ & \cr
&& $90^\circ$ && $20^\circ$, $120^\circ$, $40^\circ$ && $(1.94,1.63)$
&& $-110^\circ$, $-30^\circ$, $-40^\circ$ && $(0.41,-0.34)$ & \cr
\omit&height2pt&\omit&&\omit&&\omit&&\omit&&\omit&\cr
\tablerule }}
\caption{Same as Table \protect\ref{E1}, but with the true
$(\alpha,\beta,\gamma) = (110^\circ,30^\circ,40^\circ)$, corresponding to
$(\rho,\eta)=(0.41,0.34)$.}
\label{E3}
\end{table}

\begin{table}
\vbox{\tabskip=0pt \offinterlineskip
\def\tablerule{\noalign{\hrule}}
\halign to500pt{\strut#&
\vrule#\tabskip=1em plus 2em&
\hfil#& \vrule\,\vrule#&
\hfil#\hfil& \vrule#&
\hfil#\hfil& \vrule\,\vrule#&
\hfil#\hfil& \vrule#&
\hfil#& \vrule#\tabskip=0pt\cr\tablerule
\omit&height2pt&\omit&&\omit&&\omit&&\omit&&\omit&\cr
&&\omit\hidewidth $\theta_d$ \hidewidth&&
\omit\hidewidth $\twida_1$, $\twidb_1$, $\twidg_1$ \hidewidth&&
\omit\hidewidth  $(\rho,\eta)$ \hidewidth&&
\omit\hidewidth  $\twida_2$, $\twidb_2$, $\twidg_2$ \hidewidth&&
\omit\hidewidth $(\rho,\eta)$ \hidewidth&\cr
\omit&height2pt&\omit&&\omit&&\omit&&\omit&&\omit&\cr
\tablerule
\omit&height2pt&\omit&&\omit&&\omit&&\omit&&\omit&\cr
&& $22^\circ$ && $107^\circ$, $8^\circ$, $65^\circ$ && $(0.06,0.13)$
&& $-17^\circ$, $-98^\circ$, $-65^\circ$ && $(1.43,-3.07)$ & \cr
&& $45^\circ$ && $-50^\circ$, $-15^\circ$, $-115^\circ$ && $(-0.14,-0.31)$
&& $-40^\circ$, $-75^\circ$, $-65^\circ$ && $(0.64,-1.36)$ & \cr
&& $68^\circ$ && $-27^\circ$, $-38^\circ$, $-115^\circ$ && $(-0.57,-1.23)$
&& $-63^\circ$, $-52^\circ$, $-65^\circ$ && $(0.37,-0.80)$ & \cr
&& $90^\circ$ && $-5^\circ$, $-60^\circ$, $-115^\circ$ && $(-4.20,-9.01)$
&& $-85^\circ$, $-30^\circ$, $-65^\circ$ && $(0.21,-0.45)$ & \cr
\omit&height2pt&\omit&&\omit&&\omit&&\omit&&\omit&\cr
\tablerule }}
\caption{Same as Table \protect\ref{E1}, but with the true
$(\alpha,\beta,\gamma) = (85^\circ,30^\circ,65^\circ)$, corresponding to
$(\rho,\eta)=(0.21,0.45)$.}
\label{E4}
\end{table}

\begin{table}
\vbox{\tabskip=0pt \offinterlineskip
\def\tablerule{\noalign{\hrule}}
\halign to500pt{\strut#&
\vrule#\tabskip=1em plus 2em&
\hfil#& \vrule\,\vrule#&
\hfil#\hfil& \vrule#&
\hfil#\hfil& \vrule\,\vrule#&
\hfil#\hfil& \vrule#&
\hfil#& \vrule#\tabskip=0pt\cr\tablerule
\omit&height2pt&\omit&&\omit&&\omit&&\omit&&\omit&\cr
&&\omit\hidewidth $\theta_d$ \hidewidth&&
\omit\hidewidth $\twida_1$, $\twidb_1$, $\twidg_1$ \hidewidth&&
\omit\hidewidth  $(\rho,\eta)$ \hidewidth&&
\omit\hidewidth  $\twida_2$, $\twidb_2$, $\twidg_2$ \hidewidth&&
\omit\hidewidth $(\rho,\eta)$ \hidewidth&\cr
\omit&height2pt&\omit&&\omit&&\omit&&\omit&&\omit&\cr
\tablerule
\omit&height2pt&\omit&&\omit&&\omit&&\omit&&\omit&\cr
&& $22^\circ$ && $-45^\circ$, $-5^\circ$, $-130^\circ$ && $(-0.08,-0.09)$
&& $-45^\circ$, $-85^\circ$, $-50^\circ$ && $(0.91,-1.08)$ & \cr
&& $45^\circ$ && $-22^\circ$, $-28^\circ$, $-130^\circ$ && $(-0.81,-0.96)$
&& $-68^\circ$, $-62^\circ$, $-50^\circ$ && $(0.61,-0.73)$ & \cr
&& $68^\circ$ && $1^\circ$, $129^\circ$, $50^\circ$ && $(28.62,34.11)$
&& $-91^\circ$, $-39^\circ$, $-50^\circ$ && $(0.40,-0.48)$ & \cr
&& $90^\circ$ && $23^\circ$, $107^\circ$, $50^\circ$ && $(1.57,1.87)$
&& $-113^\circ$, $-17^\circ$, $-50^\circ$ && $(0.20,-0.24)$ & \cr
\omit&height2pt&\omit&&\omit&&\omit&&\omit&&\omit&\cr
\tablerule }}
\caption{Same as Table \protect\ref{E1}, but with the true
$(\alpha,\beta,\gamma) = (113^\circ,17^\circ,50^\circ)$, corresponding to
$(\rho,\eta)=(0.20,0.24)$.}
\label{E5}
\end{table}

Each of Tables \ref{E1}-\ref{E5} covers values of $2\theta_d$, the
angle which actually enters in $A(\bd\to\bdbar)$ [see
Eq.~(\ref{thetad})], spanning the full range $(0,\pi)$. The effect of a
negative $2\theta_d$ in the range $(-\pi,0)$ can be deduced from that
of the positive angle $2\theta_d + \pi$, since $2\theta_d$ leads to
values of $\sin 2\twida$ and $\sin 2\twidb$ opposite to those produced
by $2\theta_d + \pi$. Thus, the candidate triangle angle sets
$(\twida_1,\twidb_1,\twidg_1)$ and $(\twida_2,\twidb_2,\twidg_2)$
which correspond to a NP angle $2\theta_d$ are just the negatives of
those for $2\theta_d + \pi$.

Perusal of Tables \ref{E1}-\ref{E5} reveals several interesting
points. First, we note that most of the candidate angle sets in the
survey represented by these Tables correspond to $(\rho,\eta)$ points
well outside both the allowed $(\rho,\eta)$ region in
Fig.~\ref{rhoeta1} and its $\eta \to -\eta$ mirror image. Whenever
both candidate angle sets for a given true $(\alpha,\beta,\gamma)$ and
$\theta_d$ have $(\rho,\eta)$ values outside the allowed region,
accurate measurements of $\sin 2\twida$, $\sin 2\twidb$ and $\cos
2\twidg$ would make the presence of new physics clear. We notice,
however, that some of the candidate angle sets in Tables
\ref{E1}-\ref{E5} have $(\rho,\eta)$ values rather close to, or
actually inside, the mirror image of the allowed region. This means
that the candidates obtained when $2\theta_d$ is replaced by
$2\theta_d - \pi$ would have $(\rho, \eta)$ values inside the allowed
region itself. Whenever this happens, one would not know new physics
is present without measuring additional quantities.

As Tables \ref{E1}-\ref{E5} illustrate, if $\theta_d = {\pi \over 2}$,
the candidate angle set $(\twida_2,\twidb_2,\twidg_2)$ is always
identical to the true $(\alpha,\beta,\gamma)$, except that the
unitarity triangle has been flipped over (the common sign of the
angles has been reversed, and $\eta$ has been replaced by $-\eta$).
The reason for this is simply that when $\theta_d = \pm {\pi \over
  2}$, $\sin 2\twida = -\sin 2\alpha$ and $\sin 2\twidb = -\sin
2\beta$. Thus, since $\cos 2\twidg$ is insensitive to the sign of
$\twidg$, one of the candidate triangles looks like the real unitarity
triangle, but flipped. If one assumes that the theoretical signs of
$\BBd$ and $\BK$ are correct, the flipped character of this triangle
would imply that it cannot be the true unitarity triangle. Thus, if,
as in all of the $\theta_d = {\pi \over 2}$ examples in Tables
\ref{E1}-\ref{E5}, the other candidate triangle,
$(\twida_1,\twidb_1,\twidg_1)$, is also inconsistent with the allowed
$(\rho, \eta)$ region, one would conclude that new physics is present.
This conclusion would be confirmed by resolving the discrete
ambiguity, since as we have seen this resolution would always reject
the triangle $(\twida_2,\twidb_2,\twidg_2)$, which here is the
candidate mirroring the true unitarity triangle.

As Tables \ref{E1}-\ref{E5} also show, values of $\theta_d$ other than
${\pi \over 2}$ can also lead to candidate triangles which are at
least close to being consistent with the allowed $(\rho,\eta)$ region
of Fig.~\ref{rhoeta1} or its $\eta \to -\eta$ mirror image. One
interesting example of this phenomenon is the second row of Table
\ref{E2}, for the case
$(\alpha,\beta,\gamma)=(75^\circ,20^\circ,85^\circ)$ and $\theta_d =
45^\circ$. From this entry, it follows that $\theta_d = -45^\circ$
would lead to the candidate triangles
\beq
(\twida_1,\twidb_1,\twidg_1) = (30^\circ,65^\circ,85^\circ)
\eeq
with
\beq
(\rho,\eta) = (0.16,1.81) ~,
\eeq
and
\beq
(\twida_2,\twidb_2,\twidg_2) = (60^\circ,25^\circ,95^\circ)
\eeq
with
\beq
(\rho,\eta) = (-0.04,0.49) ~.
\eeq
This first of these is clearly inconsistent with the allowed
$(\rho,\eta)$ region of Fig.~\ref{rhoeta1}, but the second is rather
close to being consistent with it. Absent any additional information,
one would not know whether new physics is present or not. However, as
always, DAR would rule out candidate 2 -- the triangle which is close
to consistency with the Standard Model -- leaving only candidate 1,
which is completely inconsistent with the SM. Once again, new physics
would thereby be clearly established.

In fact, by changing the parameters of this example slightly, we can
easily produce a case where, despite a large $\theta_d$, one of the
candidate triangles has a $(\rho,\eta)$ actually inside the allowed
region. Suppose that $(\alpha,\beta,\gamma) =
(70^\circ,20^\circ,90^\circ)$, so that $(\rho,\eta)=(0.0,0.36)$ is
inside the allowed region, and that $\theta_d = -50^\circ$. The
measured $\sin 2\twida$, $\sin 2\twidb$ and $\cos 2\twidg$ would then
lead to the candidate triangles
\beq
\label{goodex}
(\twida_1,\twidb_1,\twidg_1) = (20^\circ,70^\circ,90^\circ)
\eeq
with
\beq
(\rho,\eta) = (0.00,2.75) ~,
\eeq
and
\beq
(\twida_2,\twidb_2,\twidg_2) = (70^\circ,20^\circ,90^\circ)
\eeq
with
\beq
(\rho,\eta) = (0.00,0.36) ~.
\eeq
Triangle 1 is completely inconsistent with the allowed $(\rho,\eta)$
region, but triangle 2 happens to coincide with the true SM unitarity
triangle (despite the presence of a large $\theta_d$!), and so is
totally consistent with the SM. As above, a DAR would select candidate
1, which is inconsistent with the allowed $(\rho,\eta)$ region. Thus,
once again, such a measurement would establish that new physics is
present.

As these examples explicitly illustrate, if one of the two candidate
triangles is fully consistent with the allowed $(\rho,\eta)$ region
and the other is far from consistent with it, the assumption that the
consistent candidate represents the true CP-violating phases in $B$
decay and no new physics is present can be completely erroneous. In
these examples the resolution of the discrete ambiguity would reveal
that the Standard Model-consistent candidate does {\it not} represent
the true CP phases, and that new physics {\it is} present.

As a final example, consider the third line of Table \ref{E5}. From
this entry, it follows that if $\theta_d$ were $-22^\circ$, the two
candidate triangles would be
\beq
(\twida_1,\twidb_1,\twidg_1) = (91^\circ,39^\circ,50^\circ)
\eeq
with
\beq
(\rho,\eta) = (0.40,0.48) ~,
\eeq
and
\beq
(\twida_2,\twidb_2,\twidg_2) = (-1^\circ,-129^\circ,-50^\circ)
\eeq
with
\beq
(\rho,\eta) = (28.62,-34.11) ~.
\eeq
Here, triangle 1 is (nearly) consistent with the allowed $(\rho,\eta)$
region, while triangle 2 is not. In this situation, DAR, by ruling out
triangle 2 as always, will select the consistent solution. This might
tempt one to erroneously conclude that there is no new physics
present. However, as this example illustrates, large new-physics
effects may actually be present in CP asymmetries even when these
asymmetries appear to be consistent with the SM. Although this appears
to contradict the conclusions of the earlier discussion, in fact this
is an example of a ``small'' $\theta_d$. That is, the true value of
$\beta$ is $17^\circ$, near the lower bound of Eq.~\ref{betac2}, while
the new-physics value is $\twidb_1 = 39^\circ$, very near the upper
bound. In other words, the new physics has essentially moved the apex
of the unitarity triangle from a point within the allowed
$(\rho,\eta)$ region to another point in the region. In such cases,
DAR will not help to uncover the new physics.  As noted before, this
is why a reduction of the allowed region would be quite helpful.

To summarize: from the expressions which determine the candidate
triangles 1 and 2 that correspond to given $(\alpha,\beta,\gamma)$ and
$\theta_d$, it is relatively straightforward to show that neither
candidate will be consistent with the presently-allowed $(\rho,\eta)$
region unless
\beq
(i)~~ 80^\circ \lsim \gamma \lsim 100^\circ
\eeq
or
\beq (ii)~~ 0 \le |\theta_d| \lsim 20^\circ ~.
\eeq
However, when one of these conditions is met, or approximately met,
then as the examples we have considered show, it is indeed possible
for a candidate triangle to be nearly or fully consistent with the
allowed region. When this occurs, one cannot establish that new
physics is present without an additional measurement. Resolution of
the discrete ambiguity, as described in Sec.~2, would often provide
the needed information. The only exception is when $\theta_d$ is close
to 0 or $\pi$.

\section{$\BBd$ and $\BK$}

It is interesting to notice that the conclusion that new physics is
present can sometimes depend crucially on the theoretical signs of the
bag parameters $\BBd$ and $\BK$ being correct. To see this, suppose
that one of the two candidate triangles implied by the measured values
of $\sin 2\twida$, $\sin 2\twidb$, and $\cos 2\twidg$ is consistent
with the allowed $(\rho,\eta)$ region or with its $\eta \to -\eta$
mirror image, while the other candidate triangle is not. Suppose
further that a DAR selects the triangle which is not consistent with
either the allowed $(\rho,\eta)$ region or its mirror image. If we
assume that the theoretical signs of $\BBd$ and $\BK$ are correct, as
we have been doing, then we can conclude that new physics is present.
Imagine, however, that we allow for the possibility that the
theoretical Sign$(\BBd)$ and/or Sign$(\BK)$ is wrong. Can we still
conclude that new physics is present?

The answer to this question is ``no''. The main point is that our
assumed techniques for determining $\sin 2\twida$ and $\sin 2\twidb$,
and for DAR, all depend on an interference between a $\bd$ decay-path
which involves $\bd - \bdbar$ mixing and one which does not. If the
Sign$(\BBd)$ used when interpreting the data is wrong, then the
assumed relation between the measured interference and the underlying
parameters will be wrong. One can show that, as a result, the
extracted $\sin 2\twida$, $\sin 2\twidb$, and any of the quantities
(such as $\cos 2\twida$) to be used by DAR, will have the wrong sign.
This means that the two candidate triangles deduced from the data will
be upside down relative to what they should be, and that the DAR will
select the wrong candidate triangle. Thus, when DAR appears to select
the candidate triangle which is not consistent with the allowed
$(\rho,\eta)$ region, this could be due to Sign$(\BBd)$ being wrong,
rather than to this triangle representing the truth and new physics
being present. The ability to conclude unambiguously that new physics
is present has been lost.\footnote{This is also true when one measures
  $\sin 2\twidg$ in $\bs(t) \to D_s^\pm K^\mp$, which involves $\bs -
  \bsbar$, rather than $\bd - \bdbar$, mixing. As we have noted, when
  $\sin 2\twida$, $\sin 2\twidb$, and $\sin 2\twidg$ are known, there
  is only one candidate triangle. However, if the signs of the bag
  parameters used to determine this triangle are wrong, it may not
  represent the true $\twida$, $\twidb$, and $\twidg$. Suppose, for
  instance, that $(\twida$, $\twidb$, $\twidg) =
  (70^\circ,30^\circ,80^\circ)$. If Sign$(\BBd)$ is right, but
  Sign$(\BBs)$ is wrong, we will get $(20^\circ,60^\circ,100^\circ)$.
  }

The situation is summarized in detail in Table~\ref{tab9}. In
constructing this table, we have assumed that one of the two candidate
triangles is consistent with the allowed $(\rho,\eta)$ region or with
its mirror image, that the other candidate triangle is not consistent
with either, and that DAR selects the latter triangle. For all
possible orientations of the candidate triangles, we indicate whether
these circumstances still imply the presence of new physics when the
theoretical sign of $\BK$ and/or $\BBd$ is wrong. Table~\ref{tab9}
makes clear that, regardless of the orientations of the candidate
triangles, if the theoretical Sign$(\BBd)$ might be wrong, then one
cannot unambiguously conclude that new physics is present. To be
sure, if the allowed-$(\rho,\eta)$-region-consistent candidate
triangle points up, then this conclusion {\em is} still possible if
one knows for certain that the theoretical Sign$(\BK)$ is correct.
However, realistically, if one is unsure of Sign$(\BBd)$, then one is
unsure of Sign$(\BK)$ as well.

\begin{table}
\begin{tabular}{|c|c|c|c|c|}  \hline
Allowed $(\rho,\eta)$ & Allowed $(\rho,\eta)$ & $\BK$ & $\BBd$ & NP 
Definitely \\
Consistent Candidate & Inconsistent Candidate &  &   & Present \\ \hline
Up              &       Up      & Right & Right         & Yes   \\
                &               & Wrong & Right         & Yes   \\
                &               & Right & Wrong         & Yes   \\
                &               & Wrong & Wrong         & No    \\  \hline
Down    &  Down & Right & Right         & Yes   \\
                &               & Wrong & Right         & Yes   \\
                &               & Right & Wrong         & No    \\
                &               & Wrong & Wrong         & Yes   \\  \hline
Up              &  Down & Right & Right         & Yes   \\
                &               & Wrong & Right         & Yes   \\
                &               & Right & Wrong         & Yes   \\
                &               & Wrong & Wrong         & No    \\  \hline
Down    &       Up      & Right & Right         & Yes   \\
                &               & Wrong & Right         & Yes   \\
                &               & Right & Wrong         & No    \\
                &               & Wrong & Wrong         & Yes   \\  \hline
\end{tabular}
\caption{The effect of incorrect signs of $\BK$ and $\BBd$ on one's
  ability to conclude that new physics (NP) is present. In the first
  column, we indicate whether the candidate triangle consistent with
  the allowed $(\rho,\eta)$ region or its mirror image points up
  $(\eta >0)$ or down $(\eta <0)$. In the second column, we show the
  same thing for the candidate triangle which is not consistent with
  either the allowed $(\rho,\eta)$ region or its mirror image. In the
  third and fourth columns, we indicate whether the theoretical signs
  of $\BK$ and $\BBd$ are right or wrong. In the final column, we
  indicate whether, under the stated assumptions concerning the signs
  of $\BK$ and $\BBd$, one would still be able to conclude that NP is
  definitely present.}
\label{tab9}
\end{table}

None of this is meant to cast doubt on the theoretically-determined
signs of $\BK$ and $\BBd$. However, since these signs are not
experimentally verified, it is important to recognize the crucial role
that they may prove to play in establishing the existence of new
physics.

\section{Conclusions}

Within the standard model, CP violation is due to complex phases in
the CKM matrix. This explanation will be tested through the
measurement of CP-violating asymmetries in the $B$ system. Such
measurements will permit the extraction of the interior angles
$\alpha$, $\beta$ and $\gamma$ of the unitarity triangle. If the
unitarity triangle constructed from these CP angles is inconsistent
with the $(\rho,\eta)$ region allowed by other measurements
($|V_{cb}|$, $|V_{ub}/V_{cb}|$, $B_d$ and $B_s$ mixing, $\epsilon_K$),
this will signal the presence of new physics.

Unfortunately, it is not the CP angles themselves which will be
measured, but rather trigonometric functions of these angles. In
particular, it is very likely that the first measurements will extract
the functions $\sin 2\alpha$, $\sin 2\beta$ and $\sin^2 \gamma$ (or
equivalently $\cos 2\gamma$). This implies that the CP angles can be
obtained only up to discrete ambiguities. In this paper we have
demonstrated that, under the assumption that the angles are the
interior angles of a triangle (i.e.\ that they are all of the same
sign and add up to $\pm \pi$), a twofold discrete ambiguity remains in
the triangle angle set $(\alpha,\beta,\gamma)$. That is, there are two
sets of solutions which form a triangle and still reproduce the
experimental results.

If one does not allow for the possibility of new physics, this
discrete ambiguity causes no problems. As shown in the paper, at most
one of these two solutions is consistent with present experimental
constraints on the unitarity triangle. Thus, in the absence of new
physics, one simply chooses that solution which is consistent with the
allowed $(\rho,\eta)$ region.

On the other hand, if one allows for the possibility of new physics,
then there may be a problem. In the presence of new physics which
modifies the phase of $\bd - \bdbar$ mixing, the CP angles measured
are not the SM angles $\alpha$, $\beta$ and $\gamma$, but rather
$\twida=\alpha + \theta_d$, $\twidb = \beta - \theta_d$ and $\twidg =
\gamma$, where $\theta_d$ is the modification due to new physics. Even
in this case, though, there are still two candidate triangle angle
sets. If both solutions are inconsistent with the allowed
$(\rho,\eta)$ region, then new physics is clearly present. But if one
solution is consistent with this region, while the other is not, then
one cannot be certain whether new physics is or is not present. In
this case it is necessary to resolve the discrete ambiguity.

We have briefly reviewed the various methods for discrete ambiguity
resolution (DAR). There is a class of techniques which use
measurements of indirect, mixing-induced CP violation to extract
different trigonometric functions of the CP angles: $\cos 2\twida$,
$\cos 2\twidb$, $\sin 2\twidg$, or $\sin 2(2\twidb + \twidg)$. Any one
of these measurements is sufficent to remove the twofold discrete
ambiguity, even in the presence of new physics. (In fact, if one
measures $\sin 2\twida$, $\sin 2\twidb$ and $\sin 2\twidg$, there is
no discrete ambiguity at all in the triangle angle set -- the
measurement of $\cos 2\twidg$ is not even necessary.) There is also a
second class of techniques which rely on direct CP violation due to
the interference of a tree and a penguin amplitude. However, these
cannot be used to reliably resolve the discrete ambiguity if new
physics is present. The reason is that new physics can affect not only
$B$ mixing, but also the penguin contributions, and these new-physics
effects may be quite different.

In this paper, we have systematically studied what kind of information
is obtained from the resolution of the discrete ambiguity when new
physics is present. We have shown that, regardless of the value of
$\theta_d$, DAR almost always reveals the presence of new physics by
choosing the solution which is inconsistent with the allowed
$(\rho,\eta)$ region. The only exception is when $\theta_d$ is near 0
or $\pi$. In addition, we have shown that, even when DAR does show
that new physics is present, it does not always choose the true values
of $(\twida,\twidb,\twidg)$.

We have explicitly demonstrated all of these points with the help of
several examples. By scanning over the parameter space, we have found
a number of different values of the SM $(\alpha,\beta,\gamma$) and
new-physics $\theta_d$ which yield the situation in which one triangle
angle set is consistent with the allowed $(\rho,\eta)$ region, while
the other is not. In most of these examples, the DAR chooses the
solution which is inconsistent with the allowed $(\rho,\eta)$ region,
thereby demonstrating that new physics is present. Without DAR, one
might have been led (erroneously) to think that the solution which is
consistent with the allowed region is the true SM solution, and no new
physics is present. The only examples in which DAR fails are those for
which $\theta_d$ is small.

Note that the analysis of our examples is strongly dependent on the
size of the allowed $(\rho,\eta)$ region. Any reduction in the allowed
region, such as an actual measurement of $B_s$ mixing or the
measurement of $B(\kl \to \pi^0 \nu \bar\nu)$, would reduce the
likelihood that one of the two candidate triangles will be consistent
with the region even when new physics is present. (Indeed, the
increasingly-stringent lower limits on $B_s$ mixing have already
helped in this regard.) If the region were sufficiently reduced, then,
except for some very fine-tuned choices of $\theta_d$, DAR would not
be necessary at all.

Finally, the above analysis is based on the assumption that the
theoretical signs of $\BBd$ and $\BK$ are correct. We have shown that
if we relax this assumption, DAR will fail. That is, if we allow
$\BBd$ and $\BK$ to be either positive or negative, then DAR cannot
definitively establish the presence of new physics. It is therefore
important to try to verify experimentally the signs of these
quantities.

\bigskip
\centerline{\bf Acknowledgments}
\bigskip
We thank L. Wolfenstein for interesting discussions. The work of DL
was financially supported by NSERC of Canada and FCAR du Qu\'ebec.

\bigskip
\noindent
{\bf Appendix A}
\medskip

In this Appendix, we present the proof that, assuming that the three
angles $\alpha$, $\beta$ and $\gamma$ are the interior angles of a
triangle, the functions $\sin 2\alpha$, $\sin 2\beta$ and $\cos
2\gamma$ determine the angle set $(\alpha,\beta,\gamma)$ up to a
twofold ambiguity.

First, suppose that $\sin 2\alpha$ and $\sin 2\beta$ have the same
sign, e.g.\ assume that they are both positive. Since $\alpha$ and
$\beta$ must have the same sign [see Eq.~(\ref{tricond1})], this implies
that $2\alpha$ and $2\beta$ both take values in the domain
$(-2\pi,-\pi)$ or $(0,\pi)$. We can immediately exclude the
$(-2\pi,-\pi)$ domain: since $|2\alpha| > \pi$ and $|2\beta| > \pi$,
this implies that $|\alpha| + |\beta| > \pi$, in violation of
Eq.~(\ref{tricond2}). Thus, for $\phi = \alpha,\beta$ we can write
\beq
2 \phi = {\pi \over 2} + 2 \delta_\phi ~~,~~~~
|\delta_\phi| < {\pi \over 4} ~.
\label{eq:2phi}
\eeq
The magnitude of $\delta_\phi$, {\it but not its sign}, is fixed by
the measured value of $\sin 2\phi$. From Eq.~(\ref{eq:2phi}) and the
assumption that $|\alpha + \beta + \gamma| = \pi$, we have
\begin{eqnarray}
\alpha & = & {\pi\over 4} + \delta_\alpha ~, \nonumber \\
\beta & = & {\pi\over 4} + \delta_\beta ~, \nonumber \\
\gamma & = & {\pi \over 2} - (\delta_\alpha + \delta_\beta) ~.
\end{eqnarray}
Now, the measured value of $\cos 2\gamma = -\cos 2(\delta_\alpha +
\delta_\beta)$ gives us the relative sign of $\delta_\alpha$ and
$\delta_\beta$. However, there still remains a twofold sign ambiguity,
corresponding to $\delta_{\alpha,\beta} \to -\delta_{\alpha,\beta}$.
This is equivalent to the twofold discrete ambiguity
\beq
\label{disamb1}
\left(\alpha, \beta, \gamma \right) \to \left({\pi \over 2} - \alpha,
{\pi \over 2} - \beta, \pi - \gamma \right).
\eeq
Note that both sets of CP angles in the above discrete ambiguity
correspond to unitarity triangles which point up.

A similar analysis holds when $\sin 2\alpha$ and $\sin 2\beta$ are
both negative, except that in this case the discrete ambiguity is
between two downward-pointing unitarity triangles:
\beq
\label{disamb2}
\left(\alpha, \beta, \gamma \right) \to \left(-{\pi \over 2} - \alpha,
-{\pi \over 2} - \beta, -\pi - \gamma \right).
\eeq

Now suppose that $\sin 2\alpha$ is positive and $\sin 2\beta$ is
negative. Thus, $2\alpha$ lies in the domain $(-2\pi,-\pi)$ or
$(0,\pi)$, while $2\beta$ is in $(-\pi,0)$ or $(\pi,2\pi)$. There are
now two cases to consider:
\begin{enumerate}

\item If $\alpha,\beta,\gamma$ are all positive, then
\begin{eqnarray}
2\alpha & = & {\pi\over 2} + 2\delta_\alpha ~,  \\
2\beta & = & {3\pi\over 2} + 2\delta_\beta ~, \label{eq:2beta}
\end{eqnarray}
where $|\delta_\alpha|,|\delta_\beta| < \pi/4$. The measured values of
$\sin 2\alpha$ and $\sin 2\beta$ determine the magnitudes of
$\delta_\alpha$ and $\delta_\beta$, but not their signs. From
Eq.~(\ref{eq:2beta}) and the assumption that $|\alpha +\beta + \gamma|
= \pi$, we have
\begin{eqnarray}
\alpha & = & {\pi\over 4} + \delta_\alpha ~, \nonumber \\
\beta & = & {3\pi\over 4} + \delta_\beta ~, \nonumber \\
\gamma & = & - (\delta_\alpha + \delta_\beta) ~.
\end{eqnarray}
The requirement that $\gamma > 0$ implies that $(\delta_\alpha +
\delta_\beta) < 0$. Now, the measured value of $\cos 2\gamma = \cos
2(\delta_\alpha + \delta_\beta)$ gives us the relative sign of
$\delta_\alpha$ and $\delta_\beta$. This, along with the constraint
that $(\delta_\alpha + \delta_\beta) < 0$, fixes $\delta_\alpha$ and
$\delta_\beta$ uniquely.

\item If $\alpha,\beta,\gamma$ are all negative, then
\begin{eqnarray}
2\alpha & = & -{3\pi\over 2} - 2\delta_\alpha ~, \nonumber \\
2\beta & = & -{\pi\over 2} - 2\delta_\beta ~,
\end{eqnarray}
with $|\delta_\alpha|,|\delta_\beta| < \pi/4$. Again, the measured values
of $\sin 2\alpha$ and $\sin 2\beta$ determine the magnitudes of
$\delta_{\alpha,\beta}$, but not their signs. We have
\begin{eqnarray}
\alpha & = & -{3\pi\over 4} - \delta_\alpha ~, \nonumber \\
\beta & = & -{\pi\over 4} - \delta_\beta ~, \nonumber \\
\gamma & = & \delta_\alpha + \delta_\beta ~.
\end{eqnarray}
This time, since $\gamma < 0$, one again requires that $(\delta_\alpha
+ \delta_\beta) < 0$. As before, the measured value of $\cos 2\gamma =
\cos 2(\delta_\alpha + \delta_\beta)$ gives us the relative sign of
$\delta_\alpha$ and $\delta_\beta$. And again, this fixes
$\delta_\alpha$ and $\delta_\beta$ uniquely once one takes into
account the constraint that $(\delta_\alpha + \delta_\beta) < 0$.

\end{enumerate}

Thus, for the case of $\sin 2\alpha >0$ and $\sin 2\beta < 0$, we have
two possible solutions for $(\alpha,\beta,\gamma)$: one with positive
values, and the other with negative values. Now, from the definitions
of $\delta_\alpha$ and $\delta_\beta$, it is clear that the magnitudes
of these quantities are the same in both cases and are determined by
the measured values of $\sin 2\alpha$ and $\sin 2\beta$. Furthermore,
for both solutions we have $(\delta_\alpha + \delta_\beta) < 0$, with
the relative sign being determined by the measurement of $\cos 2
\gamma$. Thus, the two solutions have the {\it same} values of
$\delta_\alpha$ and $\delta_\beta$. This allows us to determine the
discrete ambiguity in this case. Denoting $(\alpha,\beta,\gamma)$ as
the positive-angle solution above, the discrete ambiguity is
\beq
\label{disamb3}
\left(\alpha, \beta, \gamma \right) \to \left(-{\pi \over 2} - \alpha,
{\pi \over 2} - \beta, - \gamma \right).
\eeq
In this case, the first solution corresponds to a unitarity triangle
pointing up, while the second corresponds to one pointing down.

Finally, the analysis in the case in which $\sin 2\alpha < 0$ and
$\sin 2\beta > 0$ is clear from the above: the roles of $\alpha$ and
$\beta$ are reversed, and we have the following discrete ambiguity:
\beq
\label{disamb4}
\left(\alpha, \beta, \gamma \right) \to \left({\pi \over 2} - \alpha,
-{\pi \over 2} - \beta, - \gamma \right).
\eeq

\bigskip
\noindent
{\bf Appendix B}
\medskip

In this Appendix, we present the proof that, assuming that the three
angles $\alpha$, $\beta$ and $\gamma$ are the interior angles of a
triangle, the functions $\sin 2\alpha$, $\sin 2\beta$ and $\sin
2\gamma$ determine the angle set $(\alpha,\beta,\gamma)$ uniquely.

If $\sin 2\alpha$, $\sin 2\beta$ and $\sin 2\gamma$ are all measured,
then either (i) all three quantities are of the same sign, or (ii) two
of the three quantities are of the same sign. We consider these two
possibilities in turn.

Suppose first that $\sin 2\alpha$, $\sin 2\beta$ and $\sin 2\gamma$
are all positive. As in Appendix A, this implies that $0 <
\alpha,\beta,\gamma < \pi/2$. We can thus write
\begin{eqnarray}
\alpha & = & {\pi\over 4} + \delta_\alpha ~, \nonumber \\
\beta & = & {\pi\over 4} + \delta_\beta ~, \nonumber \\
\gamma & = & {\pi \over 2} - (\delta_\alpha + \delta_\beta) ~,
\end{eqnarray}
where $\Delta_i \equiv |\delta_i| < \pi/4$, $i=\alpha,\beta$. Since
$\gamma < \pi/2$, the bigger of $\delta_\alpha$ and $\delta_\beta$
must be positive. Suppose that it is $\delta_\alpha$. Then, depending
on the sign of $\delta_\beta$, the quantity $\delta_\alpha +
\delta_\beta$ is equal to either $\Delta_\alpha + \Delta_\beta$ or
$\Delta_\alpha - \Delta_\beta$. However, in general the measured
value of $\sin 2\gamma = \sin 2(\delta_\alpha + \delta_\beta)$ will
distinguish between these two possibilities, and so measurements of
$\sin 2\alpha$, $\sin 2\beta$ and $\sin 2\gamma$ will determine
$\alpha$, $\beta$ and $\gamma$ uniquely. The one exception is the
singular point $\Delta_\alpha = \pi/4$, or $\alpha = \pi/2$ ($\sin
2\alpha = 0$). In this case a twofold discrete ambiguity remains. A
similar analysis holds if $\Delta_\beta > \Delta_\alpha$, and also
when $\sin 2\alpha$, $\sin 2\beta$ and $\sin 2\gamma$ are all
negative.

Now suppose that two of these quantities are positive, and the third
negative, say $\sin 2\alpha,~\sin 2\beta > 0$ and $\sin 2\gamma < 0$.
This implies that $0 < \alpha,\beta < \pi/2$ and $\pi/2 < \gamma <
\pi$. Thus, we have again
\begin{eqnarray}
\alpha & = & {\pi\over 4} + \delta_\alpha ~, \nonumber \\
\beta & = & {\pi\over 4} + \delta_\beta ~, \nonumber \\
\gamma & = & {\pi \over 2} - (\delta_\alpha + \delta_\beta) ~,
\end{eqnarray}
but this time the bigger of $\delta_\alpha$ and $\delta_\beta$ must be
negative. Again, suppose that it is $\delta_\alpha$, so that
$\delta_\alpha + \delta_\beta$ is equal to either $-(\Delta_\alpha +
\Delta_\beta)$ or $-(\Delta_\alpha - \Delta_\beta)$. In general the
measured value of $\sin 2\gamma = \sin 2(\delta_\alpha +
\delta_\beta)$ will distinguish between these two possibilities, so
that measurements of $\sin 2\alpha$, $\sin 2\beta$ and $\sin 2\gamma$
will determine $\alpha$, $\beta$ and $\gamma$ uniquely. This will fail
only in the special case in which $\Delta_\alpha = \pi/4$, or $\alpha
= \pi/2$ ($\sin 2\alpha = 0$). The case $\Delta_\beta > \Delta_\alpha$
can be treated in the same way. A similar analysis holds when any two
of $\sin 2\alpha$, $\sin 2\beta$ and $\sin 2\gamma$ are of one sign,
with the third of opposite sign.


\begin{thebibliography}{99}

\bibitem{PDG} C. Caso et al.\ (Particle Data Group), \epjc{3}{98}{1}.

\bibitem{Wolfenstein} L. Wolfenstein, \prl{51}{83}{1945}.

\bibitem{AliLon} A. Ali and D. London, \epjc{9}{99}{687}.

\bibitem{NirQuinn} Y. Nir and H.R. Quinn, \prd{42}{90}{1473}.

\bibitem{signBK} Y. Grossman, B. Kayser and Y. Nir, \plb{415}{97}{90};
  I.I. Bigi and A.I. Sanda, \newprd{60}{99}{033001}.

\bibitem{CPreview} For a review, see, for example, {\it The BaBar
    Physics Book}, eds.\ P.F. Harrison and H.R. Quinn, SLAC Report
  504, October 1998.

\bibitem{CDF99} CDF Collaboration, CDF/PUB/BOTTOM/CDF/4855 (1999).

\bibitem{Dalitz} A.E. Snyder and H.R. Quinn, \prd{48}{93}{2139}.

\bibitem{BtoDK} M. Gronau and D. Wyler, \plb{265}{91}{172}. See also
  M. Gronau and D. London, \plb{253}{91}{483}; I. Dunietz,
  \plb{270}{91}{75}. Improvements to this method have recently been
  discussed by D. Atwood, I. Dunietz and A. Soni, \prl{78}{97}{3257}.

\bibitem{penguins} D. London and R. Peccei, \plb{223}{89}{257}; M.
  Gronau, \prl{63}{89}{1451}, \plb{300}{93}{163}; B. Grinstein,
  \plb{229}{89}{280}.

\bibitem{isospin} M. Gronau and D. London, \prl{65}{90}{3381}.

\bibitem{DLN} C.O. Dib, D. London and Y. Nir, \ijmp{6}{91}{1253}; M.
  Gronau and D. London, \prd{55}{97}{2845}.

\bibitem{nirsilv} Y. Nir and D. Silverman, \npb{345}{90}{301}.

\bibitem{DA} T. Goto, N. Kitazawa, Y. Okada and M. Tanaka,
  \prd{53}{96}{6662}; A.S. Dighe, I. Dunietz and R. Fleischer,
  \plb{433}{98}{147}.

\bibitem{DAWolf} L. Wolfenstein, {\it Honolulu 1997, $B$ physics and
    CP violation}, pp.\ 476-485, \prd{57}{98}{6857};

\bibitem{GNW} Y. Grossman, Y. Nir and M.P. Worah, \plb{407}{97}{307}.

\bibitem{GQ} Y. Grossman and H.R. Quinn, \prd{56}{97}{7259}.

\bibitem{ADK} R. Aleksan, I. Dunietz and B. Kayser, \zpc{54}{92}{653}.

\bibitem{Charlesetal} J. Charles, A. Le Yaouanc, L. Oliver, O. P\`ene
  and J.-C. Raynal, \plb{425}{98}{375}.

\bibitem{Azimov} Y. Azimov, \prd{42}{90}{3705}.

\bibitem{cascade} B. Kayser and L. Stodolsky, Max Planck Institute
  preprint MPI-PHT-96-112, hep-ph/9610522; B. Kayser, {\it '97
    Electroweak Interactions and Unified Theories}, ed.\ J. Tran Thanh
  Van (Editions Fronti\`eres, Paris 1997), p.\ 389.

\bibitem{IsiBs} See, for example, I. Dunietz, \prd{52}{95}{3048}, and
  references therein.

\bibitem{CPbook} {\it CP Violation}, by G.C. Branco, L. Lavoura and
J.P. Silva, Oxford University Press, 1999.

\bibitem{neutralBtoDK} M. Gronau and D. London, \plb{253}{91}{483}; I.
  Dunietz, \plb{427}{98}{179}; B. Kayser and D. London,
  hep-ph/9909561, NSF-PT-99-5, UdeM-GPP-TH-98-47.

\bibitem{NPpenguins} Y. Grossman and M.P. Worah, \plb{395}{97}{241};
  D. London and A. Soni, \plb{407}{97}{61}.

\bibitem{BurFleisch} See, for example, A.J. Buras and R. Fleischer,
  {\it Heavy Flavours II}, eds.\ A.J. Buras and M. Linder (World
  Scientific 1997), p.\ 65, and references therein.

\end{thebibliography}
\end{document}